\documentclass{mn2e}

\input{psfig.sty}

\input{times.sty}

\newcommand{\araa}{ARA\&A}   % Annual Review of Astronomy and Astrophys.
\newcommand{\aj}{AJ}         % Astronomical Journal
\newcommand{\aaa}{A\&A}      % Astronomy and Astrophysics
\newcommand{\aas}{A\&AS}     % Astronomy and Astrophys. Supplement Series
\newcommand{\aar}{A\&AR}     % Astronomy and Astrophysics Review
\newcommand{\apj}{ApJ}       % Astrophysical Journal
\newcommand{\apjs}{ApJS}     % Astrophysical Journal Supplement Series
\newcommand{\apss}{Ap\&SS}   % Astrophysics and Space Science
\newcommand{\mnras}{MNRAS}   % Monthly Notices of the Roy. Astron. Society
\newcommand{\pasp}{PASP}     % Publ. of the Astron. Society of the Pacific
\newcommand{\Msun}{\,{\rm M}_{\odot}}
\newcommand{\HST}{{\it HST}}

\newcommand{\Tza}{$T_{\it ZA93}$}

\newcommand{\HI}{H\,{\sc i}}

\def\picplace#1{\vbox{\hrule\@height 0.4pt\@width\hsize
\hbox to\hsize{\vrule\@width 0.4pt\@height#1\hfil
\vrule\@width 0.4pt\@height#1}\hrule\@height 0.4pt\@width\hsize}}
\def\squareforqed{\hbox{\rlap{$\sqcap$}$\sqcup$}}
\def\sq{\ifmmode\squareforqed\else{\unskip\nobreak\hfil
\penalty50\hskip1em\null\nobreak\hfil\squareforqed
\parfillskip=0pt\finalhyphendemerits=0\endgraf}\fi}

\def\la{\mathrel{\mathchoice {\vcenter{\offinterlineskip\halign{\hfil
$\displaystyle##$\hfil\cr<\cr\sim\cr}}}
{\vcenter{\offinterlineskip\halign{\hfil$\textstyle##$\hfil\cr
<\cr\sim\cr}}}
{\vcenter{\offinterlineskip\halign{\hfil$\scriptstyle##$\hfil\cr
<\cr\sim\cr}}}
{\vcenter{\offinterlineskip\halign{\hfil$\scriptscriptstyle##$\hfil\cr
<\cr\sim\cr}}}}}

\def\degr{\hbox{$^\circ$}}
\def\arcmin{\hbox{$^\prime$}}
\def\arcsec{\hbox{$^{\prime\prime}$}}
\def\utw{\smash{\rlap{\lower5pt\hbox{$\sim$}}}}
\def\udtw{\smash{\rlap{\lower6pt\hbox{$\approx$}}}}

\def\fs{\hbox{$.\!\!^{\rm s}$}}
\def\fdg{\hbox{$.\!\!^\circ$}}

\def\farcs{\hbox{$.\!\!^{\prime\prime}$}}
\def\hr{\hbox{$^{\rm h}$}}
\def\min{\hbox{$^{\rm m}$}}

\def\diameter{{\ifmmode\mathchoice
{\ooalign{\hfil\hbox{$\displaystyle/$}\hfil\crcr
{\hbox{$\displaystyle\mathchar"20D$}}}}
{\ooalign{\hfil\hbox{$\textstyle/$}\hfil\crcr
{\hbox{$\textstyle\mathchar"20D$}}}}
{\ooalign{\hfil\hbox{$\scriptstyle/$}\hfil\crcr
{\hbox{$\scriptstyle\mathchar"20D$}}}}
{\ooalign{\hfil\hbox{$\scriptscriptstyle/$}\hfil\crcr
{\hbox{$\scriptscriptstyle\mathchar"20D$}}}}
\else{\ooalign{\hfil/\hfil\crcr\mathhexbox20D}}%
\fi}}

\def\bbbc{{\mathchoice {\setbox0=\hbox{$\displaystyle\rm C$}\hbox{\hbox
to0pt{\kern0.4\wd0\vrule height0.9\ht0\hss}\box0}}
{\setbox0=\hbox{$\textstyle\rm C$}\hbox{\hbox
to0pt{\kern0.4\wd0\vrule height0.9\ht0\hss}\box0}}
{\setbox0=\hbox{$\scriptstyle\rm C$}\hbox{\hbox
to0pt{\kern0.4\wd0\vrule height0.9\ht0\hss}\box0}}
{\setbox0=\hbox{$\scriptscriptstyle\rm C$}\hbox{\hbox
to0pt{\kern0.4\wd0\vrule height0.9\ht0\hss}\box0}}}}
\def\bbbq{{\mathchoice {\setbox0=\hbox{$\displaystyle\rm
Q$}\hbox{\raise
0.15\ht0\hbox to0pt{\kern0.4\wd0\vrule height0.8\ht0\hss}\box0}}
{\setbox0=\hbox{$\textstyle\rm Q$}\hbox{\raise
0.15\ht0\hbox to0pt{\kern0.4\wd0\vrule height0.8\ht0\hss}\box0}}
{\setbox0=\hbox{$\scriptstyle\rm Q$}\hbox{\raise
0.15\ht0\hbox to0pt{\kern0.4\wd0\vrule height0.7\ht0\hss}\box0}}
{\setbox0=\hbox{$\scriptscriptstyle\rm Q$}\hbox{\raise
0.15\ht0\hbox to0pt{\kern0.4\wd0\vrule height0.7\ht0\hss}\box0}}}}
\def\bbbt{{\mathchoice {\setbox0=\hbox{$\displaystyle\rm
T$}\hbox{\hbox to0pt{\kern0.3\wd0\vrule height0.9\ht0\hss}\box0}}
{\setbox0=\hbox{$\textstyle\rm T$}\hbox{\hbox
to0pt{\kern0.3\wd0\vrule height0.9\ht0\hss}\box0}}
{\setbox0=\hbox{$\scriptstyle\rm T$}\hbox{\hbox
to0pt{\kern0.3\wd0\vrule height0.9\ht0\hss}\box0}}
{\setbox0=\hbox{$\scriptscriptstyle\rm T$}\hbox{\hbox
to0pt{\kern0.3\wd0\vrule height0.9\ht0\hss}\box0}}}}
\def\bbbs{{\mathchoice
{\setbox0=\hbox{$\displaystyle     \rm S$}\hbox{\raise0.5\ht0\hbox
to0pt{\kern0.35\wd0\vrule height0.45\ht0\hss}\hbox
to0pt{\kern0.55\wd0\vrule height0.5\ht0\hss}\box0}}
{\setbox0=\hbox{$\textstyle        \rm S$}\hbox{\raise0.5\ht0\hbox
to0pt{\kern0.35\wd0\vrule height0.45\ht0\hss}\hbox
to0pt{\kern0.55\wd0\vrule height0.5\ht0\hss}\box0}}
{\setbox0=\hbox{$\scriptstyle      \rm S$}\hbox{\raise0.5\ht0\hbox
to0pt{\kern0.35\wd0\vrule height0.45\ht0\hss}\raise0.05\ht0\hbox
to0pt{\kern0.5\wd0\vrule height0.45\ht0\hss}\box0}}
{\setbox0=\hbox{$\scriptscriptstyle\rm S$}\hbox{\raise0.5\ht0\hbox
to0pt{\kern0.4\wd0\vrule height0.45\ht0\hss}\raise0.05\ht0\hbox
to0pt{\kern0.55\wd0\vrule height0.45\ht0\hss}\box0}}}}
\def\bbbz{{\mathchoice {\hbox{$\sf\textstyle Z\kern-0.4em Z$}}
{\hbox{$\sf\textstyle Z\kern-0.4em Z$}}
{\hbox{$\sf\scriptstyle Z\kern-0.3em Z$}}
{\hbox{$\sf\scriptscriptstyle Z\kern-0.2em Z$}}}}

\begin{document}

\title[Globular cluster systems along the Hubble sequence of spiral galaxies]
{{\it Hubble Space Telescope\/} observations of globular cluster systems  
along the Hubble sequence of spiral galaxies}

%\addtocounter{footnote}{1}

\author[Paul Goudfrooij et al.]{
Paul Goudfrooij,$^{1}$\thanks{Electronic mail:\ goudfroo@stsci.edu} 
Jay Strader,$^{2,1}$ %\thanks{Electronic mail:\ strader@ucolick.org} 
Laura Brenneman,$^{3}$\thanks{Current Address:\ Laboratory for High
        Energy Astrophysics, Code 660, NASA Goddard Space Flight Center,
        Greenbelt, MD 20771, U.S.A.} 
Markus Kissler--Patig,$^{4}$ 
\newauthor Dante Minniti,$^{5}$ and J.\ Edwin Huizinga\,$^{1}$
%\newauthor and Dante Minniti\,$^{5}$ 
\\ 
$^1$\,Space Telescope Science Institute, 3700 San Martin Drive,
Baltimore, MD 21218, U.S.A. \\ 
$^2$\,UCO/Lick Observatory, University of California, Santa Cruz, CA
95064, U.S.A. \\
$^3$\,Astronomy Department, Williams College, 33 Lab Campus Drive,
Williamstown, MA 01267, U.S.A. \\
$^4$\,European Southern Observatory, Karl-Schwarzschild-Str.~2,
D-85748 Garching, Germany \\ 
$^5$\,Department of Astronomy, P.\ Universidad Cat\'olica, Casilla
306, Santiago 22, Chile \\ 
}

\date{Accepted 2003 April 9. Received 2003 April 8; in original form 2003 March 5}

\maketitle

\begin{abstract}
We have studied the globular cluster systems of 7 giant, 
edge-on spiral galaxies using {\it Hubble Space Telescope\/} imaging in $V$
and $I$. The galaxy sample  covers the Hubble types Sa to Sc, allowing us to
study the variation of the properties of globular cluster systems along the
Hubble sequence.  The photometry reaches $\sim$\,1.5 mag beyond the turn-over
magnitude of the globular cluster luminosity function for each galaxy.
%, allowing an accurate determination of the total number of globular clusters
% in each galaxy.   
%
Specific frequencies of globular clusters ($S_N$ values) were evaluated by
comparing the numbers of globular clusters found in our {\it WFPC2\/}
pointings with those in our Milky Way which would be detected in the
same spatial region if placed at the distance of the target
galaxies. Results from this method were found to be consistent with the more
commonly used method of constructing radial distribution functions of
globular clusters. 
The $S_N$ values of spirals with $B/T \la 0.3$ (i.e., spirals with a
Hubble type later than about Sb) are consistent with a value of $S_N = 
0.55\pm0.25$. We suggest that this population of globular clusters represents a
`universal', old halo population that is present around each galaxy. 
Most galaxies in our sample have $S_N$ values that are consistent with a
scenario in which globular cluster systems are made up of {\it (i)\/} the
aforementioned halo population plus {\it (ii)\/} a population that
is associated with bulges, which grows $\sim$\,linearly with the mass of the
bulge. Such scenarios include the `merger scenario' for the formation of
elliptical galaxies as well as the `multi-phase collapse' scenario, but it
seems inconsistent with the `secular evolution' scenario of Pfenniger \&
Norman (1990), in which bulges are formed from disc stars by means of the
redistribution of angular momentum through bar instabilities and/or
minor perturbations.   
However, there is one bulge-dominated spiral galaxy in our sample
(NGC~7814) with a low $S_N$ value that is consistent with those of the
latest-type spirals. This means that the `secular evolution' scenario
can still be viable for some bulge-dominated spirals. Thus, our
results suggest that the formation histories of galaxy bulges of
early-type spirals can be significantly different from one galaxy to another. 

\end{abstract}

\section{Introduction}
\label{s:intro}

Recent observations with the {\it Hubble Space Telescope (HST)\/} and
large-field ground-based CCD cameras have caused rapid advances in our
knowledge of the formation and evolution of globular cluster (GC) systems of 
galaxies. However, ellipticals and S0 galaxies have received by far
the largest amount of attention, mainly because early-type galaxies
usually have much richer GC systems and also suffer less from
internal extinction problems than do spirals. Consequently, our
knowledge of the GC systems in spiral galaxies is still limited to
a handful of galaxies (e.g., Harris 1991; Ashman \& Zepf 1998; Kissler-Patig et
al.\ 1999). However, it should be recognized that this
scarcity of data for spiral galaxy GC systems constitutes an important 
limitation to the use of GC systems as probes of the formation and evolution
of {\it both\/} early-type {\it and\/} late-type galaxies. 

For instance, the ability to test predictions of scenarios for the formation
of galaxies and their GCs depends on our knowledge of the `typical'
properties of GC systems of spirals.  A good example is the `merger model'
(Schweizer 1987; Ashman \& Zepf 1992), in which elliptical galaxies are
formed by mergers of spiral galaxies. In this picture, the GC systems of
elliptical galaxies are composite systems. One population of GCs is
associated with the progenitor spirals (i.e., metal-poor GCs with a halo-like
spatial distribution), while a second population of clusters forms in the
merger event (i.e., metal-rich GCs with a bulge-like spatial
distribution). Ashman and Zepf (1992; 1998) described several testable
predictions arising from this scenario, most of which are however based on
comparisons with properties of `typical' spiral galaxy GC systems.   
A particular problem in this respect is that the characteristic specific
frequency of GCs (i.e., the total number of GCs per unit galaxy luminosity)
around spirals is poorly known, especially for late-type spirals (e.g., the
compilation of Ashman \& Zepf 1998 includes specific frequency measurements
for only 5 spirals with Hubble type Sb or later). The absence of firm
constraints on the number of GCs contributed by the progenitor spirals leads 
to uncertainties in many of the key predictions for elliptical galaxy GC
systems that follow from the merger model. 
Furthermore, {\it metal-rich\/} ($-1.0 \la \mbox{[Fe/H]} \la -0.2$) GCs are
known to be associated with {\it bulges\/} of spiral galaxies like the Milky
Way and M31 rather than with their discs or halos (Minniti 1995; Barmby,
Holland \& Huchra 2001; Forbes, Brodie \& Larsen 2001), so that 
mergers of spirals containing significant bulges are likely to contribute
metal-rich GCs to the merger remnant which are {\it not\/} formed during the
merger.   It is therefore very important to establish ---using a significant
sample of spiral galaxies--- whether or not the properties of the Milky Way
(MW) GC system are typical of spiral galaxy GC systems as a class, and to
study the  relationship between bulge luminosity and the number of metal-rich
GCs in spiral galaxies.  

Another important and timely area where knowledge of the properties of GC
systems of spirals can yield significant progress is
that of the formation and evolution of bulges of spiral galaxies. 
On the one hand, structural and dynamical properties of spiral bulges 
have long shown strong similarities with those of (low-luminosity)
elliptical galaxies. Bulges follow the Fundamental Plane of elliptical
galaxies (e.g., Bender, Burstein \& Faber 1992), while their internal
dynamics are consistent with oblate, isotropic models, just like
low-luminosity ellipticals (e.g., Davies et al.\
1983). Extinction-corrected optical and near-IR colours of bulges have been
shown to be very similar to those of elliptical galaxies in the Coma cluster
(Peletier et al.\ 1999), and spectroscopic metallicities and [Mg/Fe] ratios
in bulges are similar to those of ellipticals at a given bulge luminosity
(Goudfrooij, Gorgas \& Jablonka 1999).  As ellipticals in rich clusters most
likely formed at redshift $z > 3$ (e.g., Stanford, Eisenhardt \& Dickinson
1997),  these similarities argue for an early formation of bulges, in line
with the original `monolithic collapse' model by Eggen, Lynden-Bell \&
Sandage (1962). An alternative (and quite different) model is that bulges
form from disc material through redistribution of angular momentum (Pfenniger
\& Norman 1990). In this scenario, large amounts of gas are driven into the
central region of the galaxy by a stellar bar and trigger intense star
formation. If enough mass is accreted, the bar itself will dissolve and the
resulting galaxy will reveal a  
bigger bulge than before bar formation; galaxies would thus evolve from late
to earlier types along the Hubble sequence (see also Pfenniger, Combes \&
Martinet 1994). Note that this scenario {\it does not involve GCs}, in that
only disc {\it stars\/} would contribute to the secular building of
bulges. Hence, there would be no reason for a relation 
between the bulge-to-total luminosity ratio and the number of GCs per galaxy
in this scenario. This provides a testable prediction, and is one of the key
tests performed in the present study.   

This paper is built up as follows. Section~\ref{s:approach} describes the
sample selection and the {\sl HST\/} observations, while the data analysis is
described in section~\ref{s:anal}. The various results are discussed in
section~\ref{s:disc}. Finally, section~\ref{s:concl} summarizes the main
conclusions of this study.   

\section{Approach}
\label{s:approach}

\subsection{Galaxy sample selection}
\label{s:sample}

The sample consists of spiral galaxies in an edge-on configuration. We
selected such galaxies in order to enable detection of GCs on both
sides of the spiral discs and to minimize (spatially) the impact of
dust absorption. In addition, the edge-on configuration allows one to
assess radial number density distributions of the GC systems around
these galaxies. Northern galaxies were selected from the Uppsala
General Catalogue (UGC; Nilson 1973), while southern galaxies were
selected from the Surface Photometry Catalogue of ESO-Uppsala Galaxies
(ESO-LV; Lauberts \& Valentijn 1989). Our two first-cut selection
criteria were: 
\begin{itemize}
\item the inclination of the galaxy $i \geq 80^{\circ}$, and
\item The Galactic latitude $b \geq$ 30\degr. 
\end{itemize}
Inclinations were determined following Guthrie (1992), assuming an
intrinsic flattening $q_0 \equiv (b/a)_0 = 0.11$. Inclinations were
then derived using Hubble's (1926) formula  \[ \cos^2 i = (q^2 -
q_0^2) / (1 - q_0^2)\mbox{,} \] where $q = b/a$ is the observed
(catalogued) axis ratio. 

In order to get the highest observing efficiency out of the {\sl HST\/}
observations (see below), we further considered two main (and mutually
counteracting) factors.  
On the one hand, a galaxy should be near enough to reach $M_V = -6.0$,
which is $\sim$\,1.5 magnitudes beyond the peak in the Galactic GC luminosity
function (hereafter GCLF; Harris 1996), in a few {\sl HST\/} orbits using
WFPC2. On the other hand, a galaxy should be distant enough for a 
significant fraction of its GC system to fit in the field of view of one or 
two WFPC2 exposures. 

From the remaining list of candidate spirals, we selected 2-3 galaxies of 
each main Hubble type (Sa, Sb, and Sc) to study the variation of the
properties of GC systems as a function of Hubble type (i.e., of
bulge-to-total luminosity ratio). Global properties of the final
sample galaxies are given in Table~\ref{t:sample}. 

\begin{table*}
\caption[ ]{Global properties of the sample galaxies.}
\label{t:sample}
%\begin{tabular*}{17.6cm}{@{\extracolsep{\fill}}lllllllll@{}} \hline \hline
\begin{tabular*}{17.6cm}{@{}lllllllll@{}} \hline \hline
\multicolumn{3}{c}{~~} \\ [-1.8ex]  
\multicolumn{1}{c}{Galaxy} & NGC 3628 & NGC 4013 & NGC
 4517 & NGC 4565 & NGC 4594 & IC 5176 & NGC 7814 & Reference(s) \\ [0.5ex] \hline 
\multicolumn{3}{c}{~~} \\ [-1.8ex]  
  RA (J2000) & 11\hr20\min16\fs9 & 11\hr58\min31\fs3 & 12\hr32\min45\fs6 &
     12\hr36\min20\fs8 & 12\hr39\min59\fs4 & 22\hr14\min55\fs3 &
     00\hr03\min14\fs9 & 1 \\
 DEC (J2000) & $+$13\degr35\arcmin20\arcsec & $+$43\degr56\arcmin49\arcsec &
     $+$00\degr06\arcmin48\arcsec & $+$25\degr59\arcmin16\arcsec &
     $-$11\degr37\arcmin23\arcsec & $-$66\degr50\arcmin56\arcsec &
     $+$16\degr08\arcmin44\arcsec & 1 \\
Type (RSA) & Sb  & Sb  & Sc  & Sb  & Sa  & Sbc & Sab & 2 \\
  Type (T) & 3.1 & 3.0 & 6.0 & 3.2 & 1.1 & 4.3 & 2.0 & 2 \\
$V_T$ & 9.48 & 11.23 & 10.39 & 9.58 & 8.00 & 12.12$^a$ & 10.57 & 2 \\
$V_T^0$ & 8.77 & 10.52 & 9.36 & 8.58 & 7.55 & 11.07$^a$ & 10.14 & 2$^b$ \\
$(B\!-\!V)_T$ & 0.79 & 0.96 & 0.71 & 0.84 & 1.07 & 0.90$^a$ & 0.96 & 2 \\
$i$ & 85\fdg1 & 87\fdg5 & 87\fdg4 & 86\fdg5 & 84\fdg0 & 86\fdg5 & 83\fdg0 & 10 \\
$A_V$ & 0.087 & 0.054 & 0.077 & 0.050 & 0.166 & 0.100 & 0.144 & 3$^c$ \\
$v_{\rm hel}$ & 843 & 831 & 1131 & 1229 & 1091 & 1746 & 1053 & 1 \\
$m\!-\!M$ & 29.80 & 31.35 & 31.00 & 30.06 & 29.75 & 32.16 & 30.60 &
          4,5,6,7,8,4,9$^d$ \\ 
$M_V^0$ & $-$21.03 & $-$20.83 & $-$21.64 & $-$21.48 & $-$22.20 & $-$21.09 
            & $-$20.46 & $^{b, e}$ \\
$B/T$ & 0.36 & 0.27 & 0.02 & 0.30 & 0.73 & 0.34 & 0.86 &
   10,10,11$^f$,12,13,10,14$^d$ \\ 
Disc $r_h$ & 25\farcs8 & 8\farcs4 & & 16\farcs6 & & 5\farcs4 & & 
   10,10,--,12,--,10,-- \\
Bulge $r_{\rm deV}^g$ &  & 5\farcs7 & 9\farcs9 &  & 50\farcs9 &
    & 32\farcs2 &  --,10,11,--,13,--,14$^d$ \\ 
Bulge $r_{\rm exp}$ & 22\farcs7 &  &  & 20\farcs7 &  & 3\farcs9 &  &
   10,--,--,12,--,10,--$^d$ \\  
Bulge $r_{\rm eff}$ & 38\farcs1 & 5\farcs7 & 9\farcs9 & 34\farcs7  &
   50\farcs9 & 6\farcs5 & 32\farcs2 & $^h$ \\ 
% N4594: Hes & Peletier 1993: Re = 57.1, Se = 21.03 (along minor axis) 
%      => Re, equiv = 77.7
% N4594: 48.9 along minor axis (Emsellem et al. 1994; R band) (almost the same)
% N4594, Baggett et al. 1998: s0 = 22.1, r0 = 134.2; se = 20.3, re = 50.91
%        i.e., B/D = 3.61 (50.9/134.2)**2 * -2.5 log(20.3/22.1) = 2.72
%        i.e., B/T = 0.73 (this is along major axis)
% N4594: Kent S. M., 1988, AJ 96, 514: B/T = 0.86 (bulge-dominated fit;
%        residual allocated to disk)
% Used Baggett et al. for now. 
\multicolumn{3}{c}{~~} \\ [-1.8ex] \hline 
\multicolumn{3}{c}{~~} \\ [-1.8ex] 
\end{tabular*}
 
\baselineskip=0.98\normalbaselineskip
{\small
\noindent 
%{\sl Notes to Table~\ref{t:sample}.}~~$^a$: From Lyon Extragalactic Database
{\sl Notes.}~~$^a$: From Lyon Extragalactic Database (LEDA; {\tt
  http://leda.univ-lyon1.fr}). $^b$ Corrected for both internal and
  foreground reddening. $^c$: Galactic foreground extinction only. $^d$:
  References are given in galaxy order. $^e$ Derived from the 
  above values. $^f$ Corrected value (see Section~\ref{s:S_N}.2). $^g$ All
  bulge radii are expressed in terms of the equivalent radius of an ellipse (see 
  text). $^h$ Assumed half-light radius for the bulge (see
  Section~\ref{s:S_N}.2). Equal to either $r_{\rm deV}$ or $1.678\, r_{\rm
  exp}$, depending on whether the radial light profile is fit
  better with $r^{1/4}$ or with exponential profile.
  {\sc References:} (1) NASA/IPAC Extragalactic Database (NED); (2) de
  Vaucouleurs et al.\ (1991; hereafter RC3); (3) Schlegel et al.\ (1998); (4) 
  Willick et al.\ (1997; using Tully-Fisher relation (TF)); (5) Tully \& Pierce
  (2000; using TF); (6) Ferrarese et al.\ (2000; using Cepheids); (7) Jacoby,
  Ciardullo \& Harris (1996; using a variety of methods); (8) Ford et al.\
  (1996; using Planetary Nebula Luminosity Function); (9) Tonry et al.\
  (2001; using Surface Brightness Fluctuations);  
  (10) this paper; %(10) Thronson et al.\ (1994); (10) Rice et al.\ (1996); 
  (11) Gavazzi et al.\ (2000); 
  (12) Wu et al.\ (2002); (13) Baggett, Baggett \& Anderson (1998); (14)
  Bothun, Harris \& Hesser (1992).
%(14) Jarvis \& Freeman (1985).  
}
\end{table*}

\subsection{Observations}
\label{s:obs}

\begin{figure*}
\centerline{
\psfig{figure=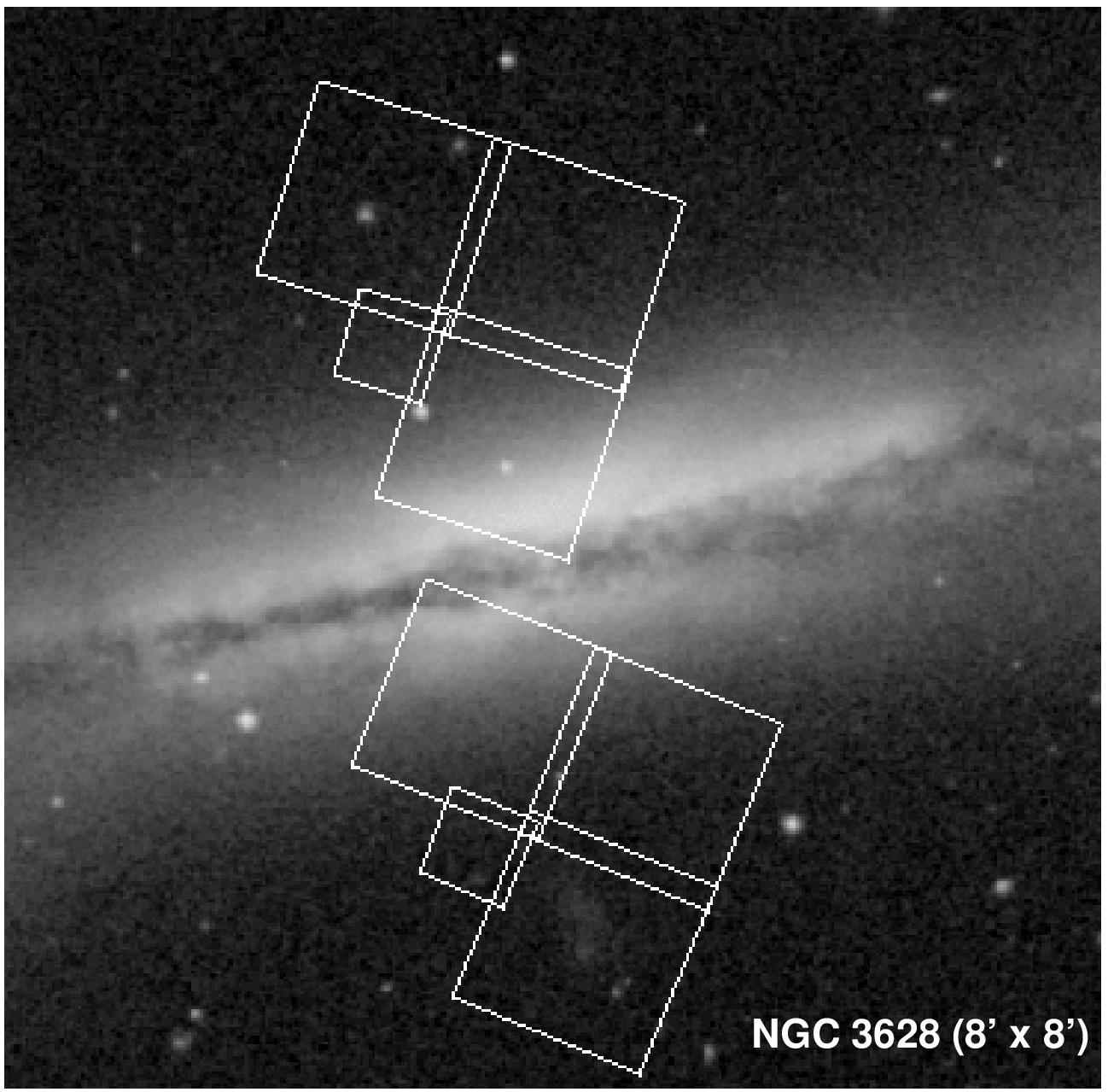,width=5.6cm,height=5.6cm}
\hspace*{1mm}
\psfig{figure=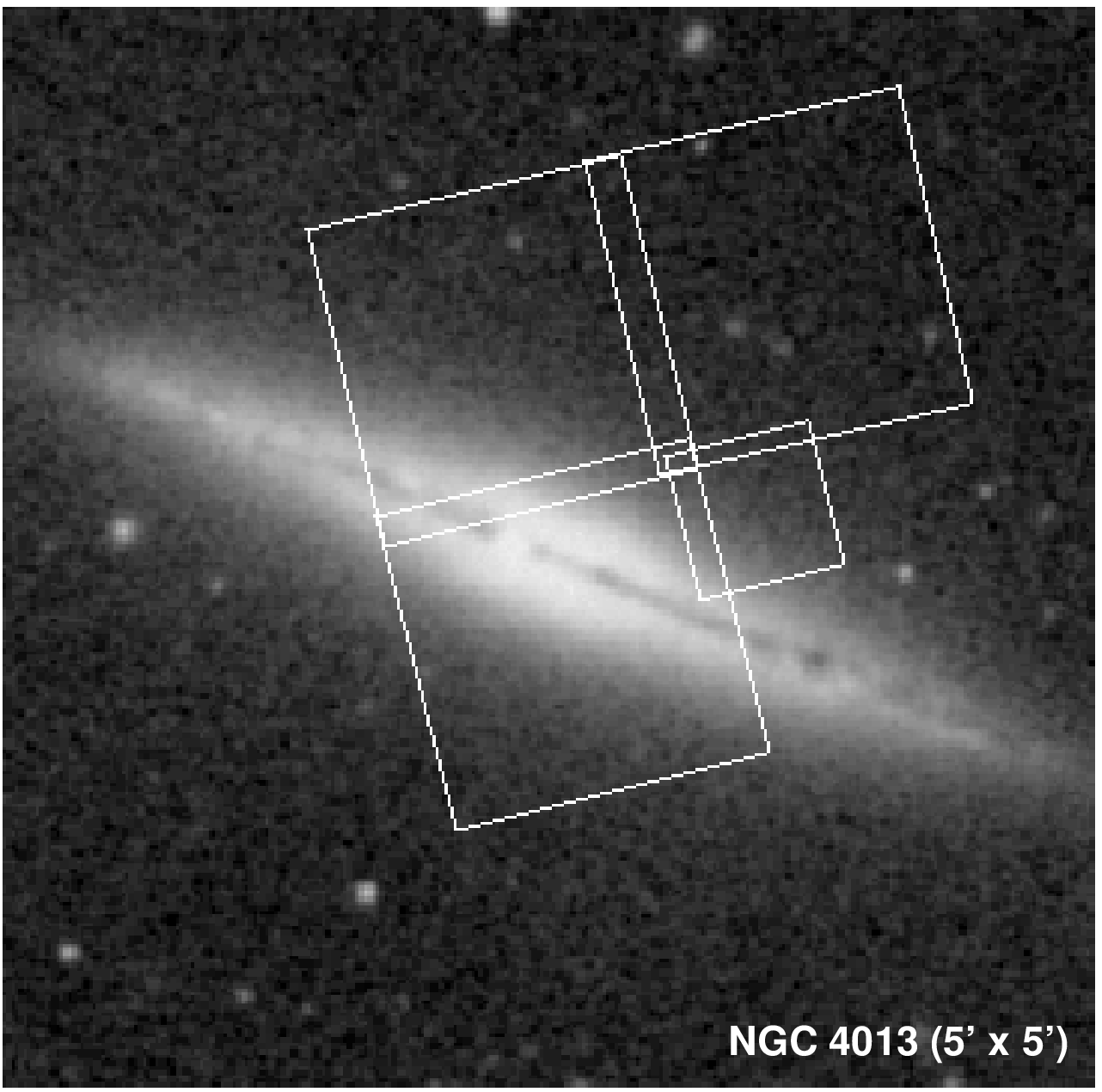,width=5.6cm,height=5.6cm}
\hspace*{1mm}
\psfig{figure=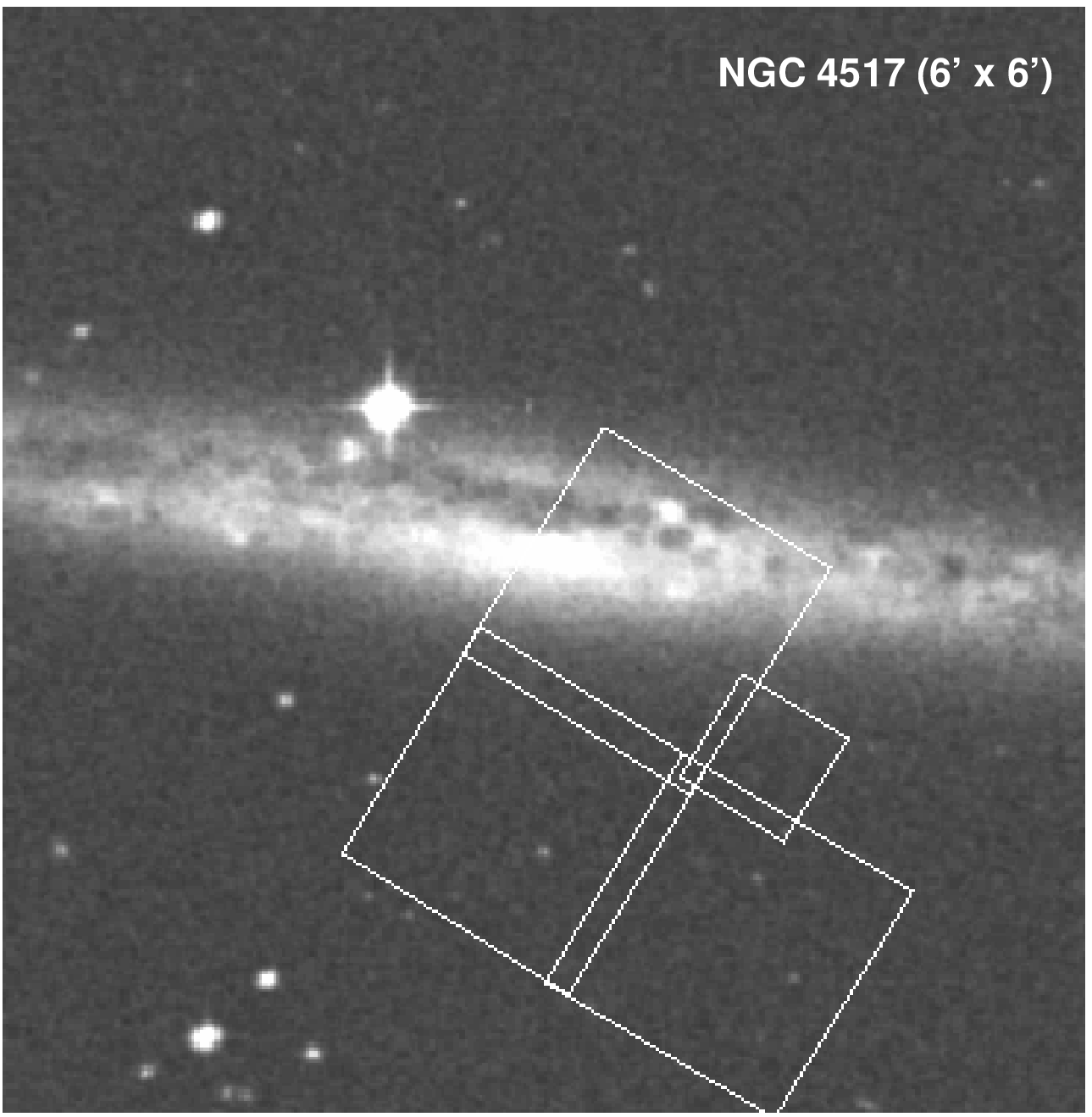,width=5.6cm,height=5.6cm}
}
\vspace*{2.5mm} 
\centerline{
\psfig{figure=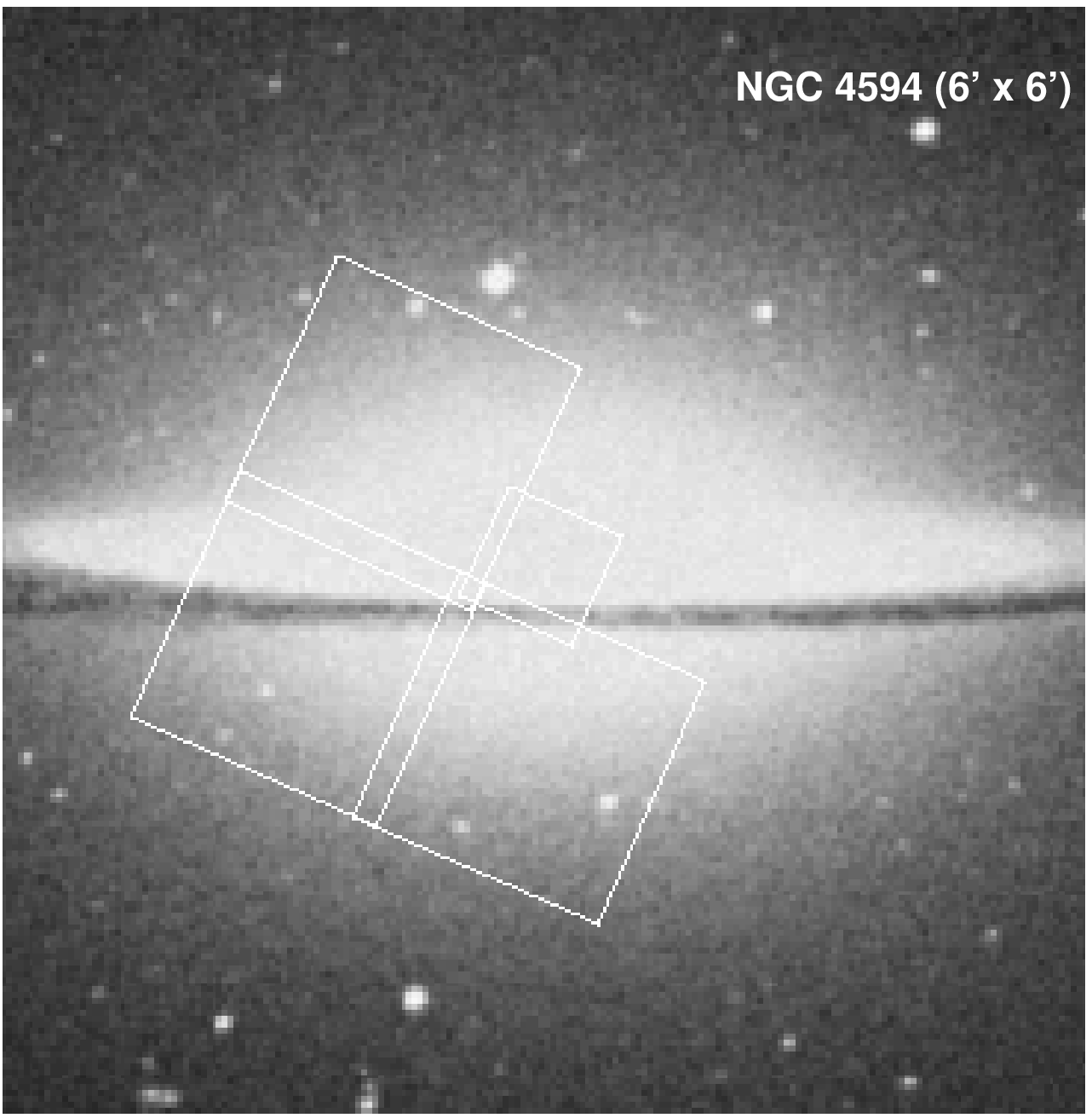,width=5.6cm,height=5.6cm}
\hspace*{1mm}
\psfig{figure=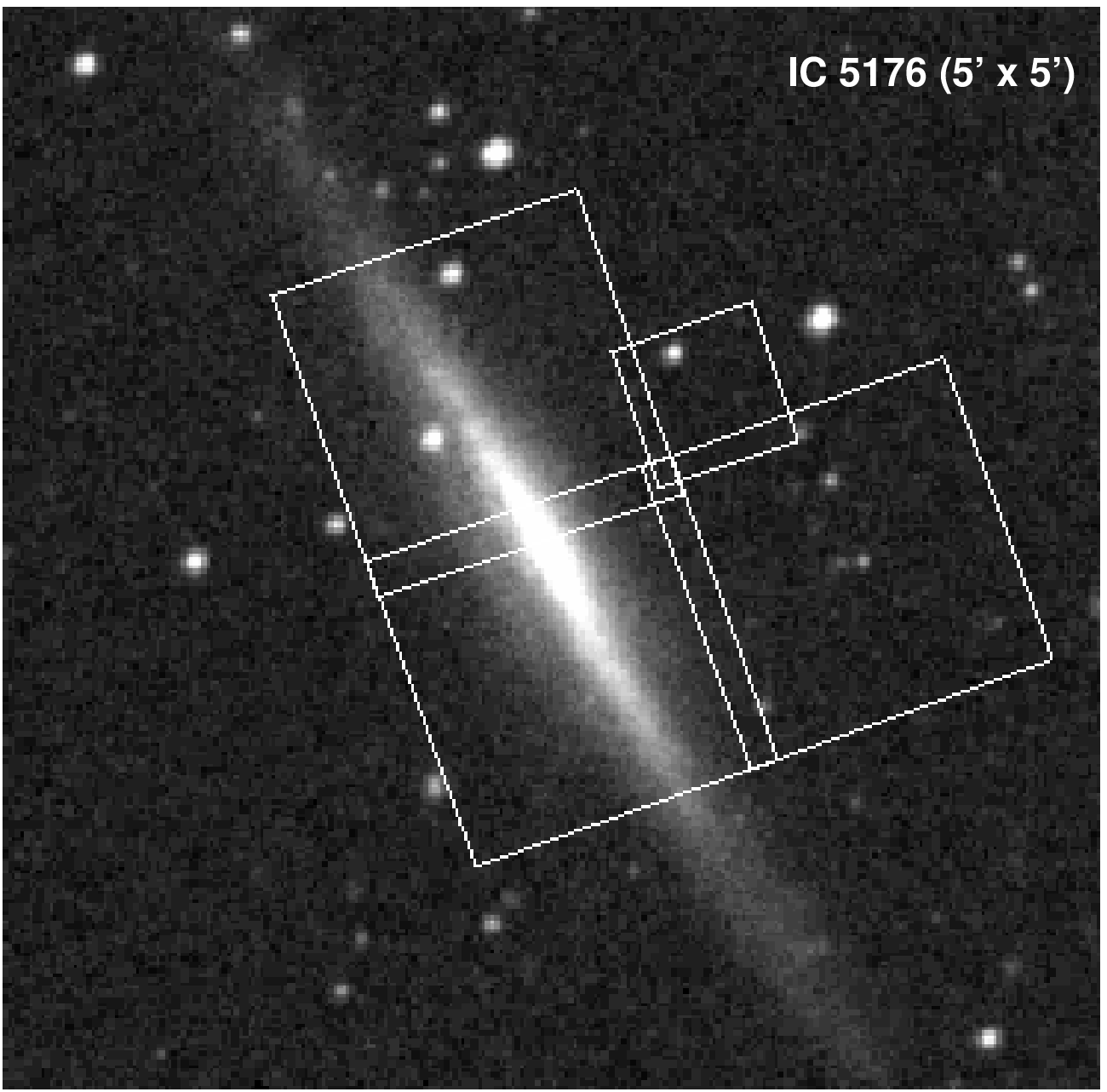,width=5.6cm,height=5.6cm}
\hspace*{1mm}
\psfig{figure=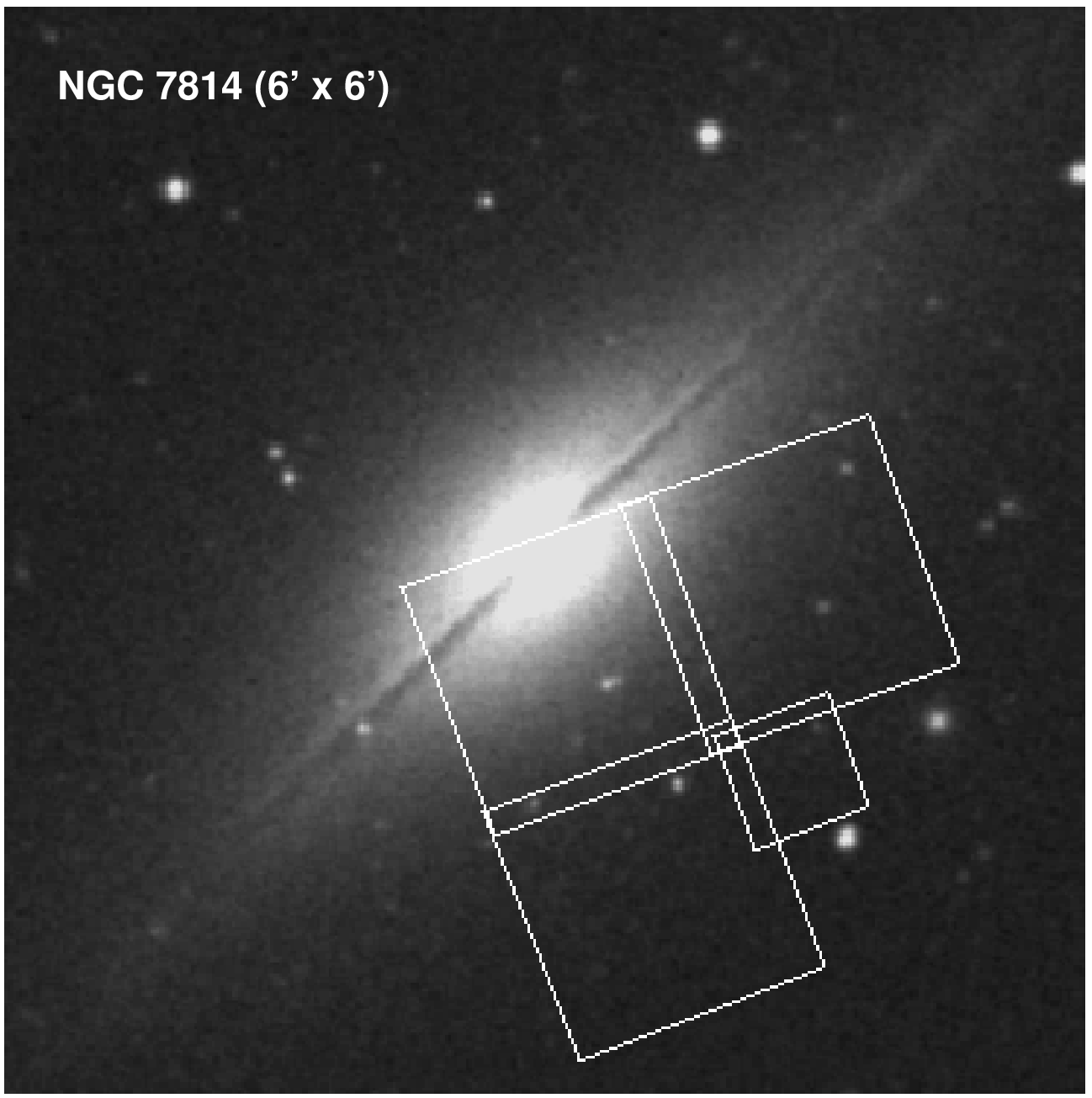,width=5.6cm,height=5.6cm}
}
\caption[ ]{Images of the sample galaxies. The greyscale images were
  extracted from the Digital Sky Survey, and the superposed lines show the
  locations of the WFPC2 field(s) of view (where each CCD chip is outlined
  individually). The galaxy identifications and field sizes are indicated in
  a corner of each image.}
\label{f:greyscales}
\end{figure*}

Observations were made with the Wide Field and Planetary Camera 2
(WFPC2) aboard {\it HST\/} as part of General Observer program 6685. The data
consist of multiple images through the F555W and F814W filters. Our program
was supplemented by archival images of a few sample galaxies (NGC~4565
and NGC~4594), taken from other {\sl HST} programs. The images of NGC~4565
were not reanalyzed; all relevant data on GC candidates in that galaxy was
taken from Kissler-Patig et al.\ (1999). 
The {\it HST\/} observations are listed in Table~\ref{t:obs} together 
with the exposure times for each galaxy. For a few galaxies, a subset of the
images were spatially offset by 0\farcs5 from the others (corresponding to an
approximately integer pixel shift in both PC and WF CCDs of the WFPC2) to
enable a good correction for hot pixels. The
locations of the WFPC2 fields are shown in Fig.~\ref{f:greyscales},
superposed onto grey-scale images from the Digital Sky Survey. 

\subsection{Data reduction and photometry}
\label{s:datared}

We used the {\sc iraf}\footnote{{\sc iraf\/} is distributed by the
National Optical Astronomy Observatories, which is operated by the
Association of Research in Astronomy, Inc., under cooperative agreement
with the National Science Foundation, U.S.A.} package for data
reduction. After standard pipeline processing and alignment of the images, we
combined the images using the IRAF/STSDAS task {\sc crrej}, thereby
effectively removing both cosmic rays and hot pixels. We also trimmed each
image to exclude the obscured regions near the pyramid edges of WFPC2,
yielding 751\,$\times$\,751 usable pixels per CCD. Prior to performing source
photometry, the strongly varying galaxy background was fitted and
subtracted. The main reason for this is to minimize errors in the photometry
due to any particular choice of object aperture and sky annulus. For the CCD
chips covering a significant part of the galaxies' bulges or dusty discs, we
applied a median filter with a 30\,$\times$\,30 pixel kernel to approximate
the galaxy background.  For the other CCDs, the smooth background gradient
was fitted by a bi-cubic spline fit. 

The photometry of star cluster candidates was carried out using the {\sc
daophot-ii} \cite{stet87} package within {\sc iraf}. The objects 
were selected by applying the {\sc daofind} task to an image prepared by
dividing the data image in question by the square root of the appropriate model
image (created as described in the previous paragraph). This procedure
ensures uniform shot noise characteristics over the whole image. We
adopted fairly tight shape constraints [$-$0.6 $<$ {\it roundness}
$<$ 0.6; 0.2 $<$ {\it sharpness} $<$ 0.9] in order to exclude extended
background galaxies and faint objects distorted by noise or
any residual bad pixels. The detection threshold was set at 4 sigma
above the residual background. Although {\sc daofind} returned with
apparent point-like detections located 
within --\,or on the edge of\,-- dust features in the galaxy discs, we
decided to exclude those from further analysis due to the difficulty in
judging the location of those sources relative to the dust features along the
line of sight. 

PSF photometry was performed using the {\sc ALLSTAR} routine. For each chip, a
PSF was constructed using, when possible, 10-15 bright, isolated GCs in the
field. Aperture corrections were calculated using growth curve analysis on a
subset of the PSF objects on each chip. One advantage of PSF photometry over
the commonly-used two-pixel aperture photometry is that it avoids the
systematic photometric offset due to underestimating aperture corrections for
bright, resolved GCs. 

\begin{table}
\caption[ ]{Summary of the {\it HST/WFPC2\/} images of the sample galaxies.}
\label{t:obs}
\begin{tabular*}{8.3cm}{@{\extracolsep{\fill}}rccrrl@{}} \hline \hline
\multicolumn{3}{c}{~~} \\ [-1.8ex]  
\multicolumn{1}{c}{Galaxy~~~~} & {\it HST\/} & Filter & \multicolumn{1}{c}{Exp.}
 & \multicolumn{1}{c}{N$_{\rm exp}^{\it a}$} & Comments \\  
       & Program     & Name   & \multicolumn{1}{c}{Time (s)}   &
 &  \\ [0.5ex] \hline 
~ \\ [-1.8ex]
NGC 3628 & 6685 & F555W &  400 & 2x4 & $^{\it b, c}$ \\
         &      & F814W &  640 & 2x4 & $^{\it b, c}$ \\
NGC 4013 & 6685 & F555W &  800 &   4 & $^{\it c}$ \\
         &      & F814W & 1100 &   2 & \\
NGC 4517 & 6685 & F555W &  400 &   4 & $^{\it c}$ \\
         &      & F814W &  640 &   4 & $^{\it c}$ \\
NGC 4565 & 6092 & F450W &  600 & 2x3 & $^{\it b, c}$ \\
         &      & F814W &  480 & 2x3 & $^{\it b, c}$ \\
%NGC 4565 & 6685 & F555W &  960 & 3x4 & $^{\it c, d}$ \\
%         &      & F814W &  960 & 3x4 & $^{\it c, d}$ \\
NGC 4594 & 5512 & F547M & 1340 &   4 & \\
         &      & F814W & 1470 &   6 & \\
 IC 5176 & 6685 & F555W &  800 &   2 & \\
         &      & F814W & 1200 &   2 & \\
NGC 7814 & 6685 & F555W &  600 &   2 & \\
         &      & F814W &  600 &   2 & \\ [0.5ex] \hline
\end{tabular*}
 
\smallskip
\baselineskip=0.98\normalbaselineskip
{\small
\noindent 
%{\sl Notes to Table~\ref{t:sample}.}~~$^a$: From Lyon Extragalactic Database
{\sl Notes.}~~{\it a}: Number of exposures; {\it b}: 2 WFPC2 pointings (exp.\
times listed are per pointing); 
% {\it c}: 3 WFPC2 pointings; {\it d}: Integer-pixel dithered data. }
{\it c}: Integer-pixel dithered data. }
\end{table}

One of the main sources of uncertainty in WFPC2 photometry (in general) is
due to charge-transfer efficiency (CTE) problems of the CCDs for which 
correction recipes are available (Whitmore, Heyer \& Casertano 1999; Dolphin
2000). We applied the Whitmore et al.\ formulae to our photometry. However,
in the present case of GC photometry on a relatively high background (due to
the diffuse galaxy light), the CTE correction typically affected the
magnitudes by less than 0.02 mag and the colours by less than 0.01 mag,
negligible for the purposes of this paper. 

The measured WFPC2 STMAG magnitudes F547M, F555W, and F814W were converted
into %ground-based 
Johnson-Cousins $V$ and $I$ magnitudes using colour terms
as derived by Holtzman et al.\ \shortcite{hol+95} and Goudfrooij et al.\
\shortcite{goud+01}. Specifically, the following conversions were used: 
\begin{eqnarray}
V-\mbox{F547M} & = & (0.009 \pm 0.002) \nonumber \\
 &  & \quad \mbox{} + (0.005 \pm 0.002)\,(\mbox{F547M}-\mbox{F814W})
 \nonumber \\
 &  & \quad \mbox{} - (0.008 \pm 0.002)\,(\mbox{F547M}-\mbox{F814W})^2 
 \nonumber \\
V-\mbox{F555W} & = & (0.643 \pm 0.001) \nonumber \\
 &  & \quad \mbox{} + (0.004 \pm 0.002)\,(\mbox{F555W}-\mbox{F814W})
 \nonumber \\ 
 &  & \quad \mbox{} - (0.015 \pm 0.002)\,(\mbox{F555W}-\mbox{F814W})^2 
 \nonumber \\
I-\mbox{F814W} & = & (-1.266 \pm 0.001) \nonumber \\
 &  & \quad \mbox{} + (0.018 \pm 0.003)\,(\mbox{F555W}-\mbox{F814W})
 \nonumber \\ 
 &  & \quad \mbox{} + (0.016 \pm 0.002)\,(\mbox{F555W}-\mbox{F814W})^2 
 \nonumber 
\end{eqnarray}

Finally, the magnitudes of the point-like sources in the sample galaxies were
corrected for Galactic foreground extinction (cf.\ Table~\ref{t:sample}). The
$A_V$ values were converted to %$A_B$ and 
$A_I$ according to the Galactic extinction law of Rieke \& Lebofsky
\shortcite{rl85}.  

\subsection{Cluster candidate selection}
\label{s:select}

To select candidate GCs, we adopted a fairly generous selection criterion for
the colour cut:\ $0.3 
\leq V\!-\!I \leq 2.0$. The lower limit was set to 0.2 mag bluer than the
bluest GC in the Milky Way, %exclude Galactic white dwarfs, 
while the upper limit was set to prevent the
exclusion of somewhat reddened objects (see the bottom right panel of
Fig.~\ref{f:CMDs}). For NGC~4594, the GCs are clearly
discernable on the colour-magnitude diagram (CMD; see below), so we were
able to set a more stringent colour cut: 
$0.5 \leq V\!-\!I \leq 1.5$. We also applied a brightness cut, eliminating GC 
candidates brighter than G1, the most luminous M31 GC ($M_{V} = -10.55$, Rich
et al.\ 1996). 

To avoid complications related to excessive reddening, we excluded candidates 
lying within the dust lanes of their parent galaxies. Since this normally
accounted for only a small fraction ($\sim 5\%$) of the total detected
cluster population, this should have minimal effects on our global
conclusions. However, we have attempted to correct this bias in the same
manner as for our incomplete spatial coverage of the target galaxies (see 
below).
%For the remainder of the analysis, the GC candidates were separated into 0.5
%mag bins.  

\begin{figure*}
\centerline{
\psfig{figure=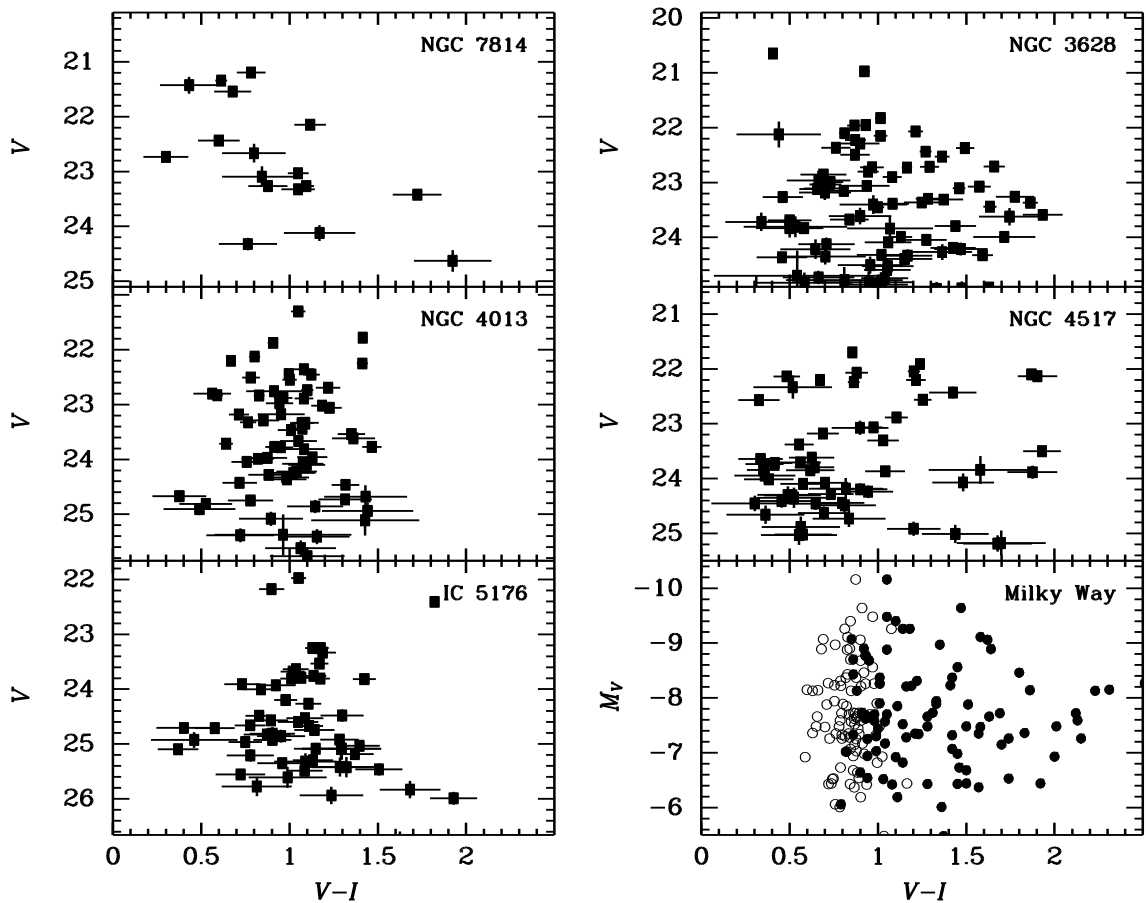,width=15cm,angle=-90.}
}
\caption[ ]{$V$ vs.\ $V\!-\!I$ colour-magnitude diagrams of candidate GCs in
  the five sample galaxies for which GC photometry was not published before,
  compared with that of the GCs in the Milky Way 
  as taken from the 2000 update of the compilation of Harris (1996; bottom
  right panel). All panels cover the same range in absolute $V$-band
  magnitudes under the assumption of the galaxy distances in
  Table~\ref{t:sample}. Filled symbols represent observed data, whereas    
  open symbols in the bottom right panel represent dereddened data of the
  Milky Way GCs. The large reddening corrections for many Milky Way GCs 
  prompted us to use fairly generous colour cuts in selecting GC candidates
  for spirals with {\sl HST\/} pointings encompassing the dusty discs (see
  Section~\ref{s:datared}). }
\label{f:CMDs}
\end{figure*}

$V$ versus $V\!-\!I$ CMDs for the candidate GCs in five of our program
galaxies\footnote{For the CMDs and the photometry tables of the GCs in
  NGC~4565 and NGC~4594, we refer to Kissler-Patig   et al.\ (1999) and
  Larsen, Forbes \& Brodie (2001), respectively} 
are shown in Fig.~\ref{f:CMDs}, together with that of the GC system of the
Milky Way as a comparison. One rather unusual finding in this context is that
the CMD of NGC~4517 reveals a relatively large number of GC candidates
with $0.3 \leq V\!-\!I \leq 0.6$, which is bluer than Galactic
GCs. However, the sizes of all these candidates (which are located on the WF
chips of WFPC2) are fully consistent with GCs at the distance of
NGC~4517. Hence we choose to retain those sources as GC candidates. 
Spectroscopy will be needed to confirm the nature of these sources. 

The photometry and positions of the 50 brightest
GC candidates in the five galaxies for which GC photometry was not published
before are listed in Tables~\ref{t:phot3628}--\ref{t:phot7814} (in
appendix~\ref{s:phottables}).    

\section{Results: Properties of the globular cluster systems}
\label{s:anal}

\subsection{The total number of globular clusters}
\label{s:NumGC}
 
The method we used to estimate the total number of GCs
in the target galaxies was to make corrections for {\it (i)\/}
contamination by Galactic foreground stars and {\it (ii)\/} spatial
coverage. The latter was done by comparing the target GC systems to that of
the Milky Way, as detailed below. 

\subsubsection{Correction for Galactic foreground stars}
First, our estimates were corrected for foreground contamination, which was
estimated from Galactic models (Bahcall \& Soneira 1981). The predicted 
star counts at the Galactic latitude and longitude of each of the sample
galaxies were subjected to the same colour and brightness cuts as the GC
candidates and then subtracted from their respective bins. The total number
of contaminating stars passing these criteria was small, generally $\sim 3-5$
stars per pointing. 

\subsubsection{Correction for completeness}
\begin{figure}
\centerline{
\psfig{figure=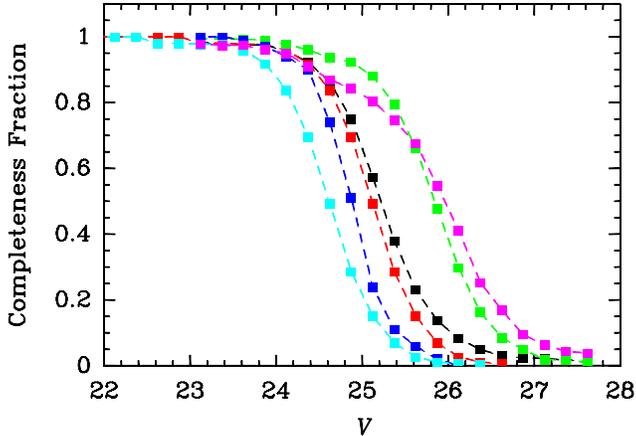,width=8.3cm,angle=-90.}
}
\caption[ ]{$V$-band completeness functions for the WFPC2 photometry of
  GC candidates. From left to right (at the 50\%
  completeness level), the curves represent the WF-chip completeness
  functions for NGC~7814, NGC~4594, NGC~4013, NGC~3628, NGC~4517, and
  IC~5176, respectively.} 
\label{f:completeness}
\end{figure}

Artificial star experiments were performed by using the {\sc daophot-ii} task
{\sc addstar} to estimate the completeness of the finding algorithm. These
were carried out in the usual fashion, using artifical GCs generated from
the PSFs constructed during the photometry measurements. These `fake' GCs
were added in groups of 100 in randomly placed positions on each chip. For
the WF chips, the 50\% completeness level was generally 1.0\,--\,1.5 mag
beyond the turnover of the GC luminosity function (GCLF), ensuring accurate
determination of the turnover point and dispersion of the GCLF. 
The WF-chip completeness functions are shown in Fig.\ 
\ref{f:completeness}. For each magnitude bin and each galaxy, a completeness
fraction was calculated, and the number of objects in the bin was divided by
this number to produce a completeness-corrected value. 
%This produced the final contamination and completeness-corrected GCLF for
%each galaxy.  

\subsubsection{Extrapolation over the luminosity function}
In preparation for the calculation of specific GC frequencies as defined by
Harris \& van den Bergh (1981), the total number of GCs is defined as twice
the number of GCs brighter than the turnover magnitude of the GCLF, where the
GCLF is assumed to be a Gaussian (in magnitude units). We therefore fitted
the completeness-corrected GCLFs by a Gaussian with a standard deviation
$\sigma = 1.1$ mag (Harris 1996; Whitmore 1997) and derived the total number
of clusters from the Gaussian fit. 

\subsubsection{Correction for incomplete spatial coverage}
To estimate the total number of GCs in each galaxy, we need to correct
for the incomplete spatial coverage of our data. Since each pointing only
covered part of the galaxy, an extrapolation over the entire galaxy was needed
to estimate the total GC population. 
% added the following sentence in the revised version
In the case of elliptical and S0 galaxies, this extrapolation is often done
by evaluating the radial distribution of GCs within the observed radial
extent of the galaxy. However, the low number statistics of GCs in most
spiral galaxies does not allow this to be done accurately. Hence the
correction for spatial coverage was done by means of a direct comparison to
the positions of GCs in the Milky Way. 
% end of extra part 
Since this process has been described and illustrated in detail in
Kissler-Patig et al.\ (1999), we will only summarize the process here.  

We created a mask defined by our spatial 
coverage of each galaxy, and applied this mask to the Milky Way. By
calculating the number of objects we would detect in the Milky Way mask, if
placed at the distance of the target galaxy, we calculated the total number
of GCs $N_{GC}$ as follows: 
%\begin{equation}
\[
        N_{GC} = N_{OBS}\,\frac{N_{MW}}{N_{MSK}} 
\]
%\end{equation}
where $N_{MSK}$ is the number of objects detected in the mask, and $N_{MW}$ is
the total number of GCs in the Milky Way. It should be noted that there are
four different orientations of the mask that would preserve the position
angle of the target galaxy (these can be seen as reflections of the mask
across its horizontal or vertical axes). Our final value for $N_{MSK}$ was
the average of the values of all four possible orientations, and the standard
deviation of the four values was taken into account in the uncertainty of
$N_{MSK}$. 

For this study, data regarding Milky Way GCs were taken from the
McMaster catalog (Harris 1996), containing 141 GCs with appropriate
($V\!-\!I$) colour information. 
%However, we adopted $N_{MW} = 180 \pm 20$ (Ashman \& Zepf 1998) as the total
%number of GCs in the Milky Way.  
However, we adopted $N_{MW} = 160 \pm 20$ (van den Bergh 1999) as the total
number of GCs in the Milky Way.  
The undetected Milky Way GCs are assumed to be behind the Galactic
bulge. Since the scale lengths and scale heights of the thin discs of the
sample galaxies are very similar to those of the Milky Way (e.g., Bothun,
Harris \& Hesser 1992; Morrison, Boroson \& Harding 1994;  Baggett, Baggett
\&  Anderson 1998; Gavazzi et al.\ 2000; Wu et al.\ 2002), the counterparts
of such GCs would not be detected in our observations either. We therefore
assume that the percentage of obscured GCs is the same in all sample
galaxies.   

We calculated a statistical uncertainty of the total number of GC candidates
calculated this way by considering Poisson errors in the observed number of
GCs, Poisson errors in the average number of Milky Way GCs in the mask
(averaged over the four different orientations), and errors in the number of
contaminating foreground objects. Our final numbers for the total population
of GC candidates are listed in Table~\ref{t:S_Ns}. 

\subsubsection{Comparison with previous studies}
% added this sentence to the text in revised version
A potential caveat of our method to correct for incomplete spatial coverage
is the implicit assumption that the spatial distribution of the GCs in the
sample galaxies is similar to that of the Milky Way globular cluster
system. Hence a comparison with other methods is in order. 
% end of extra sentence
Unfortunately, there have been only very few ground-based, wide-field studies
of spiral galaxy GC systems, which is likely due to the complications
discussed in the  Introduction. Although this is slowly changing with the use
of wide-field cameras on 4-m-class telescopes (Rhode \& Zepf 2002), we can
currently only make a useful comparison for one galaxy: NGC~7814, for which a
previous (ground-based) study of the GC system was performed by Bothun et
al.\ (1992). Bothun et al.\ found a much larger number of GC candidates than
we did ($N_{GC} = 498\pm164$ versus our $106\pm28$). From an inspection of
our WFPC2 frames, we suspect that the different result is due to their
counting method, which was based on object overdensities (relative to a
background field) rather than colours and sizes. We find a large number of
small galaxies in our WFPC2 field, which Bothun et al.\ (1992) likely counted
along with `real' GCs, while they were excluded from our GC candidate
list. This once again illustrates {\it HST}'s unique power in
discriminating GCs from small, nucleated galaxies.  

A perhaps more useful comparison of our method with others can be done for
the case of NGC~4594, for which Larsen et al.\ (2001) used the same {\it
  HST\/} data, but applied a different method: Larsen et al.\
constructed radial distribution functions of GCs to correct for incomplete
spatial coverage. While that method might in principle be expected to produce
more accurate results than the Milky Way mask method used here, it requires a
large population of GCs to avoid complications related to small number
statistics. (In our galaxy sample, this requirement is {\it only\/} met in the
case of NGC~4594). In any case, a comparison of Larsen et al.'s results with ours
is very encouraging:\ They find a total population of $N_{GC} = 1150\pm575$,
compared to $N_{GC} = 1270\pm308$ (statistical error only) in this
study. Especially heartening in this respect is that NGC~4594 (the Sombrero
galaxy) is a typical Sa galaxy, which might perhaps not be expected to
have a GC system particularly similar to that of the Milky Way, an Sbc galaxy. 

\subsection{Cluster specific frequencies}
\label{s:S_N}
 
Traditionally, the specific frequency of GCs in galaxies, $S_N$, is defined
as $S_N \equiv N_{GC} \times 10^{0.4\,(M_V+15)}$ (Harris \& van den Bergh 1981),
i.e., the total number of GCs per unit galaxy $V$-band luminosity (normalized
to $M_V = -15$). $S_N$ was introduced primarily for use in early-type
(E and S0) galaxies, where there is little variation in stellar
populations between individual galaxies. This is not necessarily the
case for spirals, and indeed $S_N$ is generally known to increase
along the Hubble sequence (going from late-- to early-type). Averaging
specific frequencies per Hubble type, the recent compilation of Ashman
\& Zepf (1998) supplemented by the study of two spiral galaxies by
Kissler-Patig et al.\ (1999) yields $\left< S_N \right> = 0.4 \pm 0.2$
for Sc spirals to $1.9 \pm 0.5$ for elliptical galaxies outside galaxy
clusters (see Section~\ref{s:disc}). Zepf \& Ashman (1993) attempted to
account for differences in stellar mass-to-light ratios among galaxy
Hubble types in a statistical sense by introducing a parameter $T$
(hereafter $T_{\it ZA93}$) to be the number of GCs per unit stellar
mass (10$^9 \,\Msun$) of a galaxy. Conversion from luminosity to mass
was achieved my assuming a characteristic $M/L_V$ value for each
galaxy Hubble type. 

\begin{table*}
\caption[ ]{Number of GCs and specific frequencies for the sample galaxies.}
\label{t:S_Ns}
\begin{tabular*}{17.2cm}{@{\extracolsep{\fill}}rcccccccc@{}} \hline \hline
\multicolumn{3}{c}{~~} \\ [-1.8ex]  
\multicolumn{1}{c}{Galaxy~~~~} & \multicolumn{1}{c}{GCs} &
 \multicolumn{1}{c}{GCs} & Total $S_N$ &  
 $T_{\it ZA93}$ & Metal-rich GCs & Bulge $S_N$ & Metal-rich GCs &
 Bulge $S_N$ \\
 & \multicolumn{1}{c}{(detected)} & \multicolumn{1}{c}{(total)} & & & 
 (all radii) & (all radii) & ($r < 
 2\, r_{1/2}$) & ($r < 2\, r_{1/2}$) \\ 
\multicolumn{1}{c}{(1)~~} & \multicolumn{1}{c}{(2)} & \multicolumn{1}{c}{(3)} &
 \multicolumn{1}{c}{(4)} & (5) & (6) & (7) & (8) & (9) \\ [0.5ex] 
\hline 
~ \\ [-1.8ex]
NGC 3628 & 92 $\pm$ 1\rlap{0} & 497 $\pm$ 11\rlap{0} & 1.9 $\pm$ 0.2 & 
 3.7 $\pm$ 1.0 & \llap{2}13 $\pm$ 48 & 0.83 $\pm$ 0.11 & ---\rlap{$^a$} 
 & --- \\ 
% & 0.73 $\pm$ 0.06 & \llap{1}5 $\pm$ 6 & 0.16 $\pm$ 0.08 \\
NGC 4013 & 69 $\pm$ 8 & 243 $\pm$ 51 & 1.1 $\pm$ 0.3 & 2.2 $\pm$ 0.7 & 
 95 $\pm$ 20 & 0.44 $\pm$ 0.18 &         3 $\pm$ 3 & 0.06 $\pm$ 0.06 \\
NGC 4517 & 62 $\pm$ 8 & 270 $\pm$ 60 & 0.6 $\pm$ 0.2 & 1.4 $\pm$ 0.5 & 
 81 $\pm$ 18 & 0.18 $\pm$ 0.16 & ---\rlap{$^b$} & --- \\
NGC 4565 & 40 $\pm$ 6 & 204 $\pm$ 38\rlap{$^c$} & 0.6 $\pm$ 0.2 & 
 1.0 $\pm$ 0.3 & \llap{1}22 $\pm$ 23 & 0.31 $\pm$ 0.15 & 
 \llap{2}0 $\pm$ 5\rlap{$^d$} & 0.17 $\pm$ 0.15 \\ 
NGC 4594 & \llap{1}59 $\pm$ 1\rlap{3} & \llap{1}270 $\pm$ 30\rlap{8} & 
 1.7 $\pm$ 0.6 & 3.6 $\pm$ 1.1 & \llap{6}91 $\pm$ 16\rlap{7} & 
 0.91 $\pm$ 0.27 & 429 $\pm$ 214 & 0.77 $\pm$ 0.44 \\
 IC 5176 & 57 $\pm$ 8 & 132 $\pm$ 25 & 0.5 $\pm$ 0.1 & 1.1 $\pm$ 0.3 & 
 67 $\pm$ 12 & 0.25 $\pm$ 0.09 & 7 $\pm$ 2\rlap{$^d$} & 0.07 $\pm$ 0.07 \\
NGC 7814 & 17 $\pm$ 4 & 106 $\pm$ 28 & 0.7 $\pm$ 0.2 & 1.5 $\pm$ 0.5 & 
 31 $\pm$ 9~~ & 0.20 $\pm$ 0.09 & 8 $\pm$ 5 & 0.06 $\pm$ 0.06 \\ [0.5ex] \hline
\end{tabular*}
 
\smallskip
\parbox{17.2cm}{
\baselineskip=0.98\normalbaselineskip
{\small
\noindent 
{\sl Notes.}~~Column (1): Galaxy name. Column (2): Detected number of GC
candidates in WFPC2 images of galaxy.  Column (3): Total number of GC
candidates around galaxy. Column (4): Total specific frequency of GC
candidates. Column (5): $T$-parameter as defined by Zepf \& Ashman
(1993) as the number of GCs per unit galaxy mass (in terms of 10$^9\;
\Msun$). Column (6): Total number of metal-rich GCs ([Fe/H] $> -1$) around
galaxy. Column (7): Total bulge specific frequency of metal-rich GC
candidates. Column (8): Number of metal-rich GC candidates within 2 bulge 
half-light radii. Column (9): Bulge specific frequency of metal-rich GC
candidates within 2 bulge half-light radii. Note {\it a}: Not measured due to
too small spatial coverage of inner bulge. Note {\it b}: Bulge 
radius too small to find bulge GCs. Note {\it c}: Taken from Kissler-Patig et
al.\ (1999). Note {\it d}: Number is twice the GC candidates found on the
non-dusty side of the nucleus.}  
}
\end{table*}

\subsubsection{Total specific frequencies}

The $S_N$ and $T_{\it ZA93}$  values of the target galaxies are listed
in Table~\ref{t:S_Ns}, 
calculated using the absolute magnitudes given in Table~\ref{t:sample} and
the number of GCs derived in Section~\ref{s:NumGC}. The quoted
uncertainties were derived by taking into account the random errors discussed
above as well as an uncertainty of 0.2 mag in the (total) absolute,
dereddened $V$-band magnitudes of the 
galaxies. $T_{\it ZA93}$ values were calculated by converting the
galaxy luminosities to masses following Zepf \& Ashman (1993),
using $M/L_V$ values of 5.4, 6.1, and 5.0 for 
Sa, Sab-Sb, and Sbc-Sc galaxies, respectively (cf.\ Faber \& Gallagher
1979). We note that the uncertainty in the {\it distance\/} of a
galaxy has a negligible effect on the $S_N$ and $T_{\it ZA93}$ values, since
the change in the total number of GCs (due to a change in the physical
area covered by the WFPC2 frames) is compensated by a similar change
of the luminosity of the galaxy.  

\subsubsection{Bulge specific frequencies}

As already mentioned in the Introduction, a view is emerging that inner 
metal-rich GCs in spiral galaxies may be associated with their bulges
rather than with their (thick) discs. Ample evidence for this is available
for the Milky Way and 
M\,31, where the metal-rich GC systems show kinematics, metallicities, and
spatial distributions matching those of the underlying bulge stars (Minniti
1995; C\^ot\'e 1999; Barmby et al.\ 2001; Perrett et al.\ 2002). Such a 
physical association is also likely present for elliptical galaxies, where
the spatial distribution of metal-rich GCs typically follows that of the
spheroidal galaxy light distribution very closely, whereas the metal-poor GC
system typically has a more extended distribution (e.g., Ashman \& Zepf 1998
and references therein). 
If indeed this association can be confirmed for a large sample of 
spirals as well, it would provide an important causal link between the
formation of a spheroidal stellar system and that of metal-rich GCs. A
suggestion that this might be the case was recently provided by Forbes et
al.\ (2001) who compared the GC systems of the Milky Way (an Sbc galaxy),
M\,31 (Sb), and M\,104 (Sa). They argued that the ``bulge specific
frequency (bulge $S_N$)'' -- which they defined as the number of metal-rich GCs
within 2 bulge effective radii from the galaxy centres divided by the bulge
luminosity (normalized to $M_V = -15$) -- was consistent among those three
galaxies (bulge $S_N \sim 0.5$), and similar to typical values for field
elliptical galaxies as well.  

Structural bulge properties and bulge-to-total luminosity ratios
(hereafter $B/T$ ratios) for the
sample galaxies are listed in Table~\ref{t:sample}. For 
NGC~4517, NGC~4565, NGC~4594 and NGC~7814, these numbers were taken from the
literature (Gavazzi et al.\ 2000; Wu et al.\ 2002; Baggett, Baggett \&
Anderson 1998; Bothun, Harris \& Hesser 1992, respectively). The bulge/disc
decompositions for the other sample 
galaxies are described in Appendix \ref{s:bulgedisc}. For NGC~4013 and
IC~5176, we used our own WFPC2 images (in the F555W band). For NGC~3628, we
used the WIYN $V$-band image published by Howk \& Savage (1999) 
which was graciously made available to us.  
Among the sample galaxies, only the bulges of NGC~4013, NGC~4517, NGC~4594
and NGC~7814 were well fit by a de Vaucouleurs' profile. The
bulges of NGC~3628, NGC~4565 and IC~5176 were much better fit by an
exponential profile. Half-light radii $r_{1/2}$ for the bulges of the
latter galaxies were calculated as $1.678 \, r_0$, where $r_0$ is the
exponential scale length of the profile. The final half-light bulge radii
$r_{1/2}$ of all sample galaxies are listed in the last row of
Table~\ref{t:sample}.  

\begin{figure*}
\centerline{
\psfig{figure=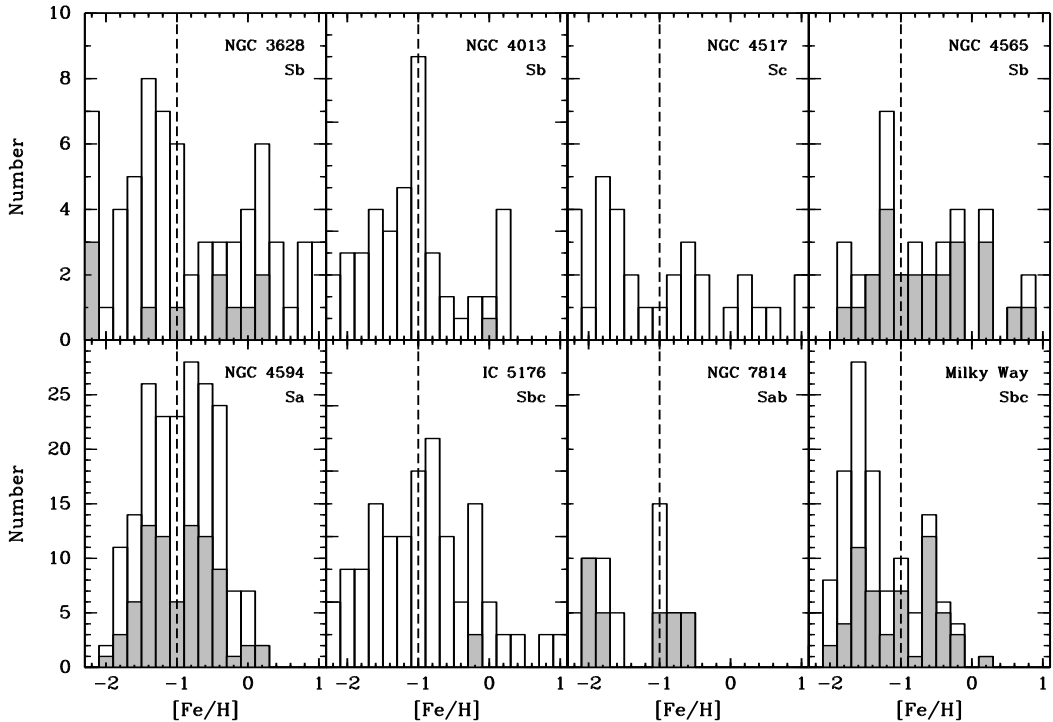,width=14cm,angle=-90.}
}
\caption[ ]{Metallicity distributions of GC candidates in the spiral galaxies
  in our sample, and of GCs in the Milky Way for comparison. The $y$-axis
  scaling is different for each galaxy, but the small tickmarks always
  represent 1 GC. Open histograms represent the total observed GC system, and
  hatched histograms represent observed GCs within 2 bulge half-light
  radii from the galactic center. The vertical dashed line at [Fe/H] =
  $-$1 separates `metal-poor' from `metal-rich' GC candidates.}  
\label{f:histograms}
\end{figure*}

To identify `metal-rich' GCs in our sample galaxies, we
considered a lower limit of [Fe/H] = $-$1, which is the location of the `dip'
in the metallicity distribution of the Milky Way GC system (Harris
1996). This is shown in Fig.~\ref{f:histograms}, superposed onto the
GC metallicity histograms of the sample galaxies. The metallicities were
derived by transforming the $V\!-\!I$ colours (corrected for Galactic
extinction) into [Fe/H] values using the recent calibration of Kissler-Patig
et al.\ (1998). The $B\!-\!I$ colours of GCs in NGC~4565 were converted into
[Fe/H] values using the relation for Milky Way GCs, given in Couture, Harris
\& Allwright (1990). 
%a least-squares fit of [Fe/H] vs.\ $(B\!-\!I)_0$ as listed in
%the Milky Way GC database of Harris (1996). The resulting relation is 
%\[
%\mbox{[Fe/H]} = (2.26 \pm 0.25) \times (B\!-\!I)_0 - (4.88 \pm 0.39) 
%\]
%which has an RMS uncertainty of 0.38 dex. 
Note that the [Fe/H] values for GC candidates in external spiral
galaxies as derived from colours are likely overestimates given the
possibility of reddening by dust, especially in the inner regions. Thus, the
number of ``metal-rich'' clusters resulting from this exercise has to be
regarded as an upper limit. 

To allow a direct comparison with the `bulge $S_N$' values defined in
Forbes et al.\ (2001), we counted the total number of `metal-rich' GC
candidates as well as those with a projected galactocentric radius $< 2 \,
r_{1/2}$ identified in our images. Correction for incomplete spatial coverage
was done by dividing the observed number of GCs by the fraction of the total
area within $2\, r_{1/2}$  sampled by our images. Since the surface density
of GCs typically falls off with increasing galactocentric radius, this
correction procedure was done in elliptical annuli, i.e., multiplying the
number of GCs in each annulus by the ratio of total annulus area to that
which was actually observed. We chose to skip this measurement for the case of
NGC~3628 since the \HST\ observations of NGC~3628 only covered a very small
portion of the inner bulge, rendering the extrapolated result very
uncertain. In the cases of NGC~4565 and IC~5176, we only counted GCs on one
side of the disc for this purpose, namely the side that was visually not
significantly impacted by dust extinction.  The bulge $S_N$ values for the
sample galaxies were then calculated using the $B/T$ ratios listed in
Table~\ref{t:sample}. The results are listed in Table~\ref{t:S_Ns}. 

The bulge $S_N$ values within $2\, r_{1/2}$ for most of our galaxies 
are roughly consistent with a value of $\sim$\,0.1, with the exception
of NGC~4594, the Sombrero galaxy, for which we find a value of
$\sim$\,0.5. At face value, the low bulge $S_N$ values for most of our
galaxies seem to be inconsistent with the prediction of Forbes et
al.\ (2001) who suggested that bulge $S_N$ values within $2\, r_{1/2}$
are constant among spiral galaxies with a value around 0.5. However,
systematic uncertainties of the inner bulge $S_N$ values as derived
from our data can be substantial. E.g., we generally optimized our
observing strategy to detect GCs in the halos of the target galaxies
(cf.\ Fig.~\ref{f:greyscales}), and dust extinction effects are strong
in the innermost bulge regions. Hence we defer further discussion on the
universality of the bulge $S_N$ until high-resolution near-IR data are
available. 
% A good approach to this would be a dedicated observing program using
% near-IR data (to minimize the influence of extinction in the dusty
% inner regions). 
Our following discussion on trends of $S_N$ values along the Hubble
sequence will therefore focus on the `total' $S_N$ values. 

\section{Discussion}
\label{s:disc}

%\subsection{Relationships with Hubble type}
In the context of the `major merger' (Ashman \& Zepf 1992) or `multi-phase
collapse' (Forbes, Brodie \& Grillmair 1997)  scenarios for the formation of
early-type galaxies, the higher $S_N$ values of ellipticals (with
respect to spirals) are due to `extra' GCs that were formed with high
efficiency during the event that also formed the spherical stellar component
(bulge). If this is indeed how bulges of spirals form in general, one would
expect $S_N$ to vary systematically along the Hubble sequence of spirals
(increasing from late-type to early-type spirals). If instead bulges were
formed from disc stars by means of redistribution of angular momentum
through minor perturbations and/or bar-like instabilities as in the scenario
of Pfenniger \& Norman (1990), one would {\it not\/} expect to see any
significant variation along the Hubble sequence, since no GCs would
be formed in this secular building of bulges. 

\begin{figure*}
\centerline{
\psfig{figure=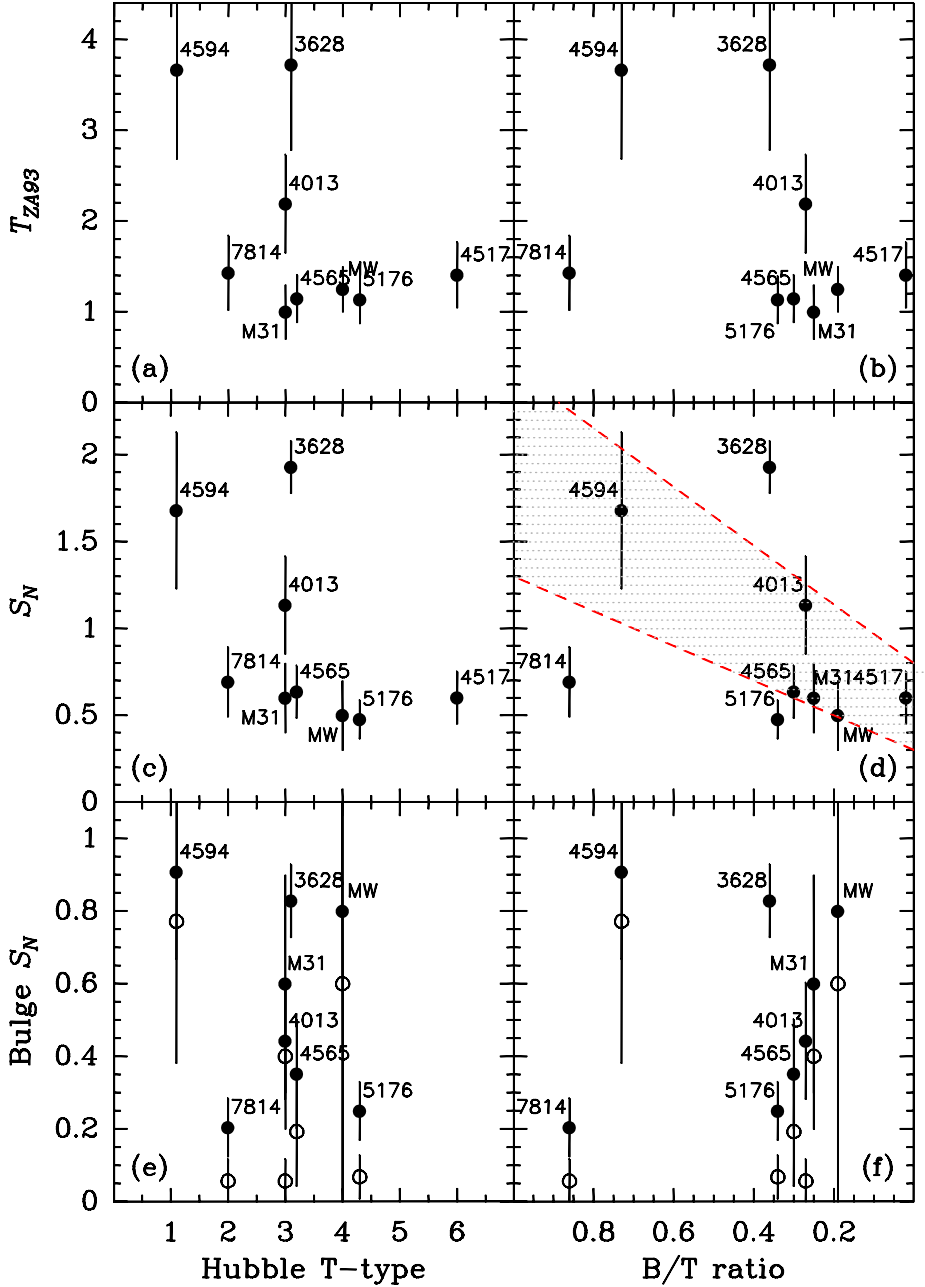,width=12cm}
}
\caption[ ]{Relationships of the specific frequency parameters ($S_N$
  and bulge $S_N$) and the $T$ parameter
  (introduced by Zepf \& Ashman 1993; hence we identify it as
  $T_{\it ZA93}$) with galaxy Hubble type and bulge-to-total luminosity
  ratio for the sample galaxies as well as the Milky Way and M31 (data
  for the latter two galaxies were taken from Forbes et al.\ 2001). The
  identifications of the galaxies are plotted next to the accompanying data
  points (at a location least impacting the other symbols). 
  The hatched area in panel (d) represents the region of expected $S_N$ values
  for a simple model described in Section~\ref{s:disc}. 
  In panels (e) and (f) (i.e., the bulge $S_N$ plots), the filled circles
  represent the `total' bulge $S_N$ values, while the open circles represent
  the bulge $S_N$ values within 2 half-light radii of the bulges (see
  Section~\ref{s:S_N}.2). To avoid overcrowding, the galaxy identifications
 are only plotted along with the filled circles.}  
\label{f:S_N_vs_type}
\end{figure*}

Since our galaxy sample spans a large range of Hubble types (from Sa to Sc),
we are in a good position to test this. We plot the $S_N$ and $T_{\it
  ZA93}$ values versus Hubble type and $B/T$ ratio in
Figure~\ref{f:S_N_vs_type}. Values for the Milky Way and M31 (taken from
Forbes et al.\ 2001 and references therein) are also included in that
Figure. There are a number of remarkable aspects of these results, which we
discuss in order below.  

%  MW: Sn = 0.5 +/- 0.2; bulge Sn = 0.8 +/- 0.9; < 2 Reff = 0.6 +/- 0.6
% M31: Sn = 0.6 +/- 0.2; bulge Sn = 0.6 +/- 0.3; < 2 Reff = 0.4 +/- 0.2

First of all, there does not seem to be any significant difference between
trends derived from \Tza\ and those from $S_N$. Hence, we will only discuss
trends involving $S_N$, with the understanding that the trends involving
\Tza\ are consistent with them.  

A general trend in Figure~\ref{f:S_N_vs_type} is that $S_N$ stays
basically constant at a value of $\sim$\,0.55 for spirals with $B/T \la 0.3$
(roughly corresponding to Hubble types Sb and later). As these galaxies are
clearly dominated in mass by their halos, this constant $S_N$ (and \Tza)
suggests that the formation process of massive (late-type) spiral galaxies
involves the creation of a quite constant number of GCs per unit galaxy
luminosity (and mass). We suggest that this population of GCs represents a 
`universal', old halo population that is present around each galaxy. 

It is now well established that elliptical galaxies in environments similar to
that of the spirals in our sample (i.e., outside galaxy clusters) have higher
specific frequencies than these late-type spirals. A $\chi^2$-weighted average
$S_N$ value for ellipticals outside galaxy clusters published in the
literature is $S_N$ = 1.9 $\pm$ 0.6 (using $S_N$ data from Ashman \& Zepf
1998; Brown et al.\ 2000; Georgakakis, Forbes \& Brodie 2001; Goudfrooij et
al.\ 2001).  If indeed the higher values of $S_N$ for elliptical galaxies are
due to an `extra' population of GCs which was formed during the formation
process of spheroidal stellar systems (such as bulges) involving a rapid
dissipative collapse in which the physical conditions are such that giant star
clusters are formed very efficiently (i.e., as in the `merger' and
`multi-phase collapse' scenarios mentioned above), then one would naively
expect the $S_N$ values to scale linearly with $B/T$ ratio from $S_N =
0.55\pm0.25$ at $B/T = 0$ to $S_N = 1.9\pm0.6$ at $B/T = 1$. This area is
hatched in the middle right panel of Fig.~\ref{f:S_N_vs_type}. 

%, except perhaps for NGC~4013 (for which $S_N = 1.0 \pm 0.3$). 
Interestingly, the $S_N$ values of most sample galaxies are indeed consistent
with this simple prediction. There are however two galaxies for which this
is clearly not the case: NGC~3628 (Sb) and NGC~7814 (Sab).

The result for NGC~3628 is quite remarkable. This Sb galaxy turns out to
have a $S_N$ value %(`total' as well as `bulge' $S_N$'s) 
that is significantly higher than all other Sb or Sbc galaxies with $S_N$
measurements to date. However, NGC~3628 is a peculiar galaxy. It is 
part of the Leo Triplet (Arp 317; Arp 1966). Large-area optical
observations reveal a tidal plume extending $\sim$\,100 kpc towards the
east, as well as a bridge between NGC~3628 and NGC~3627 (Burkhead \& Hutter
1981; Chromey et al.\ 1998), both of which contain large amounts of \HI\ gas
($5.4 \times 10^8\; \Msun$, which is $\sim$\, 15\% of the total \HI\ mass
observed in the main body of NGC~3628; Haynes, Giovanelli \& Roberts
1979). Rots (1978) constructed three-body orbital models for the tidal
interactions between NGC~3627 and NGC~3628 which reproduce the formation of
the plume and bridge, yielding an age for the plume of $8 \times 10^8$ yr
since perigalacticon. This age was confirmed by photometric measurements
of bright clumps in the plume by Chromey et al.\ (1998).  It is therefore quite
conceivable that the high $S_N$ value of NGC~3628 may be due to an `extra'
supply of star clusters which were formed during the gaseous interaction that
also caused the tidal plume and bridge. This situation seems to
have occurred in M82, another nearby starburst galaxy which has undergone a
past starburst triggered by tidal interactions with a gas-rich 
neighbour, during which a population of GCs did indeed form (see, e.g., de
Grijs, O'Connell \& Gallagher 2001). As the gas from 
which such second-generation star clusters are formed is expected to be of
relatively high metallicity (relative to halo GCs), this picture is
consistent with the observation that the (total) bulge $S_N$ of NGC~3628
(i.e., the number of {\it metal-rich\/} GC candidates per unit bulge
luminosity) is higher than those of the other Sb--Sc galaxies in our sample
by a factor similar to the total $S_N$. This idea should be verified by
performing wide-field imaging as well as spectroscopy of a significant number
of GC candidates in NGC~3628. 

%{\bf What is the S_N of M82? Cannot find it anywhere}

As to the deviant result for NGC~7814, it is perhaps somewhat surprising to
find that the two earliest-type spirals in this sample (NGC~4594 and
NGC~7814) have very different specific frequencies. NGC~4594 (Sa) has $S_N$
values that are similar to those of field ellipticals (e.g., Ashman 
\& Zepf 1998) and S0 galaxies (Kundu \& Whitmore 2001), and thus seems to
have formed GCs at a similarly high efficiency relative to its field
stars. It is therefore highly unlikely that bulges like the one of NGC~4594
can be formed through the `secular evolution' scenario of Pfenniger \& Norman 
(1990). On the other hand, NGC~7814 has a much lower $S_N$, which is
consistent with the values found for the bulk of the latest-type
spirals. This difference in $S_N$ persists when considering the `bulge
$S_N$' values. 
At face value, this result suggests that the (star--) formation
histories of galaxy bulges of early-type spirals can be significantly
different from one galaxy to another. The formation process of some bulges
(such as that of NGC~4594) may have involved dissipative collapse, being
accompanied by the creation of a significant number of (metal-rich) GCs,
while bulges in other spiral galaxies (like NGC~7814) may have been built
from the redistribution of angular momentum through bar-like instabilities
as in the `secular evolution' scenario.  
In this context, it may well be relevant that NGC~7814 is the least luminous
(and thus presumably least massive) spiral galaxy in our sample, being
$\sim$\,5 times less luminous than NGC~4594. As Pfenniger (1991) showed, the
accretion of small satellite galaxies (with a mass of $\sim$\,10\% of that of
the massive galaxy) can trigger the formation of a bar and produce a bulge
through secular evolution. It is tempting to suggest that the difference in
$S_N$ value between NGC~4594 and NGC~7814 is simply due to the fact
that satellite galaxies at a satellite-to-giant galaxy mass ratio of
$\sim$10\% are more common at low (giant) galaxy luminosities than at high
luminosities. Thus suggestion is certainly consistent with observed
luminosity functions of galaxies in poor groups (Jerjen, Binggeli \& Freeman
2000; Flint, Bolte \& Mendes de Oliveira 2001). 

% N4594: M/L = 3.6;   N7814: M/L = 8.5 (Jarvis \& Freeman 1985 ApJ 295 324)
Given the importance of this issue in the context
of our understanding of the formation of the Hubble sequence of galaxies, we
argue that this suggestion be tested with a statistically significant sample
of (edge-on) Sa\,--\,Sab spirals which covers an appropriate range of
luminosities. This can now be realized 
with only very modest amounts of observing time using the recent ACS 
camera aboard \HST, which reaches some 2 magnitudes deeper than WFPC2 
at a given exposure time. The current generation of large-area CCD mosaics on
ground-based 4-m-class telescopes will be very valuable as well in terms of
covering the whole GC populations around such galaxies in one shot. 

\section{Summary}
\label{s:concl}

We have studied the GC systems of 7 giant, edge-on spiral galaxies
covering the Hubble types Sa to Sc, allowing us to study the variation of the
properties of GC systems along the Hubble sequence. Our high-resolution {\it
  HST/WFPC2\/} data (supplemented by archival {\it WFPC2\/} data for two
galaxies) reached $\sim$\,1.5 mag beyond the turn-over magnitude 
of the GC luminosity function for each galaxy. 

Specific frequencies of GCs ($S_N$ values) were evaluated by comparing the
numbers of GCs found in our {\it   WFPC2\/} pointings with the number of GCs in
our Milky Way which would be detected in the same spatial region if placed at
the distance of the target galaxies. Results from this method were compared
with the more commonly used method of constructing radial distribution
functions of GCs to correct for incomplete spatial coverage in the case of
NGC~4594 (where both methods were possible), and found to be consistent with
one another to within 1$\sigma$. 

The $S_N$ values of spirals with $B/T \la 0.3$ (i.e., spirals with a
Hubble type later than about Sb) are consistent with a value of $S_N = 
0.55\pm0.25$. We suggest that this population of GCs represents a `universal',
old halo population that is present around each galaxy. 

Most galaxies in our sample have $S_N$ values that are consistent with a
scenario in which GC systems are made up of {\it (i)\/} the aforementioned
halo population plus {\it (ii)\/} a population that is associated
with bulges of spirals, which grows linearly with the luminosity (and mass)
of the bulge. Such scenarios include the `merger scenario' for the formation
of elliptical galaxies as well as the `multi-phase collapse' scenario, but it
seems inconsistent with the `secular evolution'  scenario of Pfenniger \&
Norman (1990), in which bulges are formed from disc stars by means of the 
redistribution of angular momentum through minor perturbations and/or bar
instabilities.  

On the other hand, the bulge-dominated spiral NGC~7814 shows a low $S_N$
value, consistent with those of the latest-type spirals. NGC~7814 is the
least luminous galaxy in our sample. Based on observed
luminosity functions of galaxies in poor groups, we suggest that the 
`secular evolution' scenario to build bulges in early-type spirals is most
viable for low-luminosity spirals. 

Thus, our results suggest that the formation histories of galaxy
bulges of early-type spirals can be significantly different from one galaxy
to another. Given the importance of our understanding of the formation of
the Hubble sequence of galaxies in the context of galaxy evolution, we argue
that the GC systems of a statistically significant sample of luminous,
edge-on Sa\,--\,Sab spirals be studied in the near future.   

\paragraph*{Acknowledgments.} \ \\ 
This paper is based on observations obtained with the NASA/ESA
{\it Hubble Space Telescope}, which is operated by AURA, Inc., under
NASA contract NAS 5--26555. 
We are grateful to Chris Howk for making his WIYN images of our Northern 
targets available to us, and thank David Clements for his contribution during
the proposal preparation stage. We thank the referee, Duncan Forbes,
for a fast, thorough and constructive review. 
We have made use of the NASA/IPAC Extragalactic Database 
(NED) which is operated by the Jet Propulsion Laboratory, Caltech, 
under contract with the National Aeronautics and Space Administration.
PG was affiliated with the Astrophysics Division of the Space Science
Department of the European Space Agency during part of this project. 
DM is supported by FONDAP Center for Astrophysics 15010003. JS and LB
would like to thank the Space Telescope Science Institute for
financial support through its Summer Student Program.

\appendix

\section{Bulge/disc decompositions}
\label{s:bulgedisc}

Since bulge/disc decompositions are not the main topic of this paper, we will 
only provide a brief description of our method here. A more detailed
description will be provided elsewhere. We first determined the scale
heights of the discs at 4 independent positions along their major axes, 
outside the region where the bulge contribution dominates. In general, the
positions of the galaxy planes were determined by assuming symmetrical light
distributions with respect to the disc plane, and the vertical profiles on
either side of the plane were averaged together. However, the latter was not
done for the bulge fits on NGC~4013 and IC~5176, since the WFPC2 frames only
covered a large enough spatial extent on one side of the galaxy centre. 

The vertical brightness distributions of the discs were well fitted by an
exponential profile 
\begin{equation}
\Sigma_{\rm D} (z) = \Sigma_{{\rm D},0} \exp (-z/z_0)\mbox{,} 
\end{equation}
where $z_0$ is the disc scale height. We found that as long as one stays
outside the central areas where there is an obvious contribution from bulge
light, the disc scale heights generally do not exhibit any significant
variation with distance along the major axis. This is consistent with the
findings of de Grijs, Peletier \& van der Kruit (1997). The average disc
scale heights are listed in Table~\ref{t:sample}.

To obtain structural parameters for the bulges of these galaxies, we 
used a least-square program that fits a two-component (disc+bulge) model to
the minor-axis brightness profiles. The disc component was fit by an  
exponential profile and the bulge component was fitted by two different
functions, namely {\it (i)\/} a de Vaucouleurs' 
(1953) profile  
\begin{equation}
\Sigma_{\rm B} (z) = \Sigma_{{\rm B,e}} \; \exp(-7.688((z/z_{\rm
e})^{1/4} - 1))\mbox{,} 
\end{equation}
where $z_{\rm e}$ is the effective radius along the minor axis and
$\Sigma_{{\rm B,e}}$ is the intensity at $z = z_{\rm e}$, and {\it (ii)\/}
a (second) exponential profile. For each galaxy, the disc scale height $z_0$  
as found above was fixed in the fit. Areas where dust extinction is prominent
(such as the central dust lanes) were flagged and ignored during the fitting
process. The choice between the de Vaucouleurs' profile and the (second)
exponential profile was dictated by the $\chi^2$ value of the fit. 

Among the sample galaxies, only the bulges of NGC~4013, NGC~4517, NGC~4594
and NGC~7814 were well fit by a de Vaucouleurs' profile. The bulges of the
other galaxies were better fit by an exponential profile. The effective radii
(or exponential scale radii r$_{\rm exp}$) of the bulges are listed in 
Table~\ref{t:sample}, after converting them to equivalent radii $r
\equiv \sqrt{ab} = b/\sqrt{1-\epsilon}$, where $a$ and $b$ are
the semimajor and semiminor axes of the ellipse, and $\epsilon$ its
ellipticity. For NGC~3628, NGC~4013, NGC~4517 and IC~5176, the bulge
ellipticities were determined by using the ellipse fitting program in the
{\sc isophote} package within IRAF. The dusty areas of the galaxies were
flagged and ignored in the ellipse fitting process.  For the galaxies with
exponential bulge profiles, we also list the bulge half-light radii (= 1.678
r$_{\rm exp}$) for comparison with the effective radii of the bulges with 
de Vaucouleurs' profiles. 

\section{Tables of globular cluster candidates}
\label{s:phottables}
The photometry and astrometry of the 50 brightest GC 
candidates on the WFPC2 images of the galaxies for which GC photometry
was not published before are given in Tables~\ref{t:phot3628} --
\ref{t:phot7814}. For the photometry of GC candidates in NGC~4565 and
NGC~4594 we refer to Kissler-Patig et al.\ (1999) and Larsen et al.\
(2001), respectively. 

\begin{table}
\caption[ ]{Photometry and astrometry of the 50 brightest globular cluster
  candidates on the WFPC2 frames of NGC~3628. The object list is sorted on
  $V$ magnitude (brightest first).}
\label{t:phot3628}
\begin{tabular*}{8.38cm}{@{}lrrcc@{}} \hline \hline
\multicolumn{3}{c}{~~} \\ [-1.8ex]  
\multicolumn{1}{c}{ID} & \multicolumn{1}{c}{RA} & \multicolumn{1}{c}{DEC} &
 $V$ & $V\!-\!I$ \\ 
  & \multicolumn{1}{c}{(J2000)} & \multicolumn{1}{c}{(J2000)} & mag & mag 
  \\ [0.5ex] \hline 
\multicolumn{3}{c}{~~} \\ [-1.8ex]  
 3628--1 & 11:20:14.64 & +13:32:27.6 & $20.65 \pm 0.01$ & $0.41 \pm 0.02$ \\ 
 3628--2 & 11:20:21.12 & +13:35:38.4 & $20.97 \pm 0.02$ & $0.92 \pm 0.03$ \\ 
 3628--3 & 11:20:19.92 & +13:35:45.6 & $21.83 \pm 0.02$ & $1.01 \pm 0.03$ \\ 
 3628--4 & 11:20:19.44 & +13:36:50.4 & $21.95 \pm 0.01$ & $0.93 \pm 0.03$ \\ 
 3628--5 & 11:20:14.40 & +13:33:57.6 & $21.96 \pm 0.03$ & $0.87 \pm 0.04$ 
 \\ [0.8ex]
 3628--6 & 11:20:17.76 & +13:35:45.6 & $22.07 \pm 0.03$ & $1.21 \pm 0.04$ \\ 
 3628--7 & 11:20:17.76 & +13:37:22.8 & $22.10 \pm 0.02$ & $0.81 \pm 0.03$ \\ 
 3628--8 & 11:20:12.72 & +13:34:04.8 & $22.12 \pm 0.24$ & $0.44 \pm 0.24$ \\ 
 3628--9 & 11:20:15.60 & +13:33:32.4 & $22.15 \pm 0.02$ & $1.01 \pm 0.04$ \\ 
3628--10 & 11:20:16.56 & +13:33:36.0 & $22.23 \pm 0.02$ & $0.87 \pm 0.06$ 
 \\ [0.8ex]
3628--11 & 11:20:12.72 & +13:33:57.6 & $22.29 \pm 0.08$ & $0.90 \pm 0.11$ \\ 
3628--12 & 11:20:19.92 & +13:35:31.2 & $22.37 \pm 0.05$ & $0.76 \pm 0.08$ \\ 
3628--13 & 11:20:16.80 & +13:35:34.8 & $22.37 \pm 0.05$ & $1.49 \pm 0.05$ \\ 
3628--14 & 11:20:16.32 & +13:35:38.4 & $22.44 \pm 0.03$ & $1.27 \pm 0.03$ \\ 
3628--15 & 11:20:18.00 & +13:35:31.2 & $22.50 \pm 0.07$ & $0.87 \pm 0.08$ 
 \\ [0.8ex]
3628--16 & 11:20:16.80 & +13:35:52.8 & $22.53 \pm 0.04$ & $1.36 \pm 0.04$ \\ 
3628--17 & 11:20:17.04 & +13:35:31.2 & $22.71 \pm 0.06$ & $1.66 \pm 0.06$ \\ 
3628--18 & 11:20:15.84 & +13:35:38.4 & $22.71 \pm 0.04$ & $1.29 \pm 0.05$ \\ 
3628--19 & 11:20:15.84 & +13:35:34.8 & $22.72 \pm 0.04$ & $0.97 \pm 0.07$ \\ 
3628--20 & 11:20:18.48 & +13:36:07.2 & $22.73 \pm 0.02$ & $1.16 \pm 0.03$ 
 \\ [0.8ex]
3628--21 & 11:20:17.76 & +13:35:45.6 & $22.80 \pm 0.04$ & $0.94 \pm 0.06$ \\ 
3628--22 & 11:20:16.08 & +13:35:34.8 & $22.86 \pm 0.10$ & $0.69 \pm 0.13$ \\ 
3628--23 & 11:20:17.28 & +13:35:42.0 & $22.90 \pm 0.04$ & $1.08 \pm 0.05$ \\ 
3628--24 & 11:20:13.44 & +13:34:04.8 & $22.96 \pm 0.17$ & $0.66 \pm 0.18$ \\ 
3628--25 & 11:20:16.08 & +13:35:38.4 & $22.99 \pm 0.05$ & $0.73 \pm 0.07$ 
 \\ [0.8ex]
3628--26 & 11:20:20.40 & +13:34:40.8 & $23.04 \pm 0.05$ & $0.70 \pm 0.09$ \\ 
3628--27 & 11:20:17.28 & +13:34:44.4 & $23.06 \pm 0.11$ & $0.94 \pm 0.13$ \\ 
3628--28 & 11:20:18.24 & +13:35:27.6 & $23.08 \pm 0.06$ & $1.57 \pm 0.06$ \\ 
3628--29 & 11:20:16.80 & +13:36:36.0 & $23.11 \pm 0.02$ & $1.46 \pm 0.04$ \\ 
3628--30 & 11:20:16.32 & +13:35:31.2 & $23.11 \pm 0.06$ & $0.68 \pm 0.10$ 
 \\ [0.8ex]
3628--31 & 11:20:17.28 & +13:35:38.4 & $23.12 \pm 0.07$ & $0.66 \pm 0.10$ \\ 
3628--32 & 11:20:19.20 & +13:35:34.8 & $23.15 \pm 0.09$ & $0.81 \pm 0.11$ \\ 
3628--33 & 11:20:19.92 & +13:35:52.8 & $23.18 \pm 0.14$ & $0.70 \pm 0.15$ \\ 
3628--34 & 11:20:18.00 & +13:35:27.6 & $23.26 \pm 0.10$ & $1.78 \pm 0.11$ \\ 
3628--35 & 11:20:20.40 & +13:35:34.8 & $23.26 \pm 0.08$ & $0.46 \pm 0.11$ 
 \\ [0.8ex]
3628--36 & 11:20:16.56 & +13:35:42.0 & $23.30 \pm 0.06$ & $1.28 \pm 0.07$ \\ 
3628--37 & 11:20:17.04 & +13:35:27.6 & $23.31 \pm 0.08$ & $1.37 \pm 0.11$ \\ 
3628--38 & 11:20:16.56 & +13:35:34.8 & $23.37 \pm 0.08$ & $1.25 \pm 0.10$ \\ 
3628--39 & 11:20:20.40 & +13:37:37.2 & $23.37 \pm 0.03$ & $1.86 \pm 0.04$ \\ 
3628--40 & 11:20:16.32 & +13:35:31.2 & $23.39 \pm 0.06$ & $1.08 \pm 0.09$
 \\ [0.8ex]
3628--41 & 11:20:18.96 & +13:35:27.6 & $23.40 \pm 0.17$ & $0.97 \pm 0.19$ \\ 
3628--42 & 11:20:18.72 & +13:34:33.6 & $23.44 \pm 0.03$ & $1.63 \pm 0.04$ \\ 
3628--43 & 11:20:18.48 & +13:33:43.2 & $23.45 \pm 0.03$ & $1.00 \pm 0.04$ \\ 
3628--44 & 11:20:17.04 & +13:35:27.6 & $23.59 \pm 0.10$ & $1.93 \pm 0.11$ \\ 
3628--45 & 11:20:18.48 & +13:36:25.2 & $23.61 \pm 0.15$ & $0.90 \pm 0.18$
 \\ [0.8ex]
3628--46 & 11:20:17.04 & +13:35:31.2 & $23.62 \pm 0.16$ & $1.75 \pm 0.17$ \\ 
3628--47 & 11:20:17.52 & +13:35:34.8 & $23.68 \pm 0.08$ & $0.84 \pm 0.11$ \\ 
3628--48 & 11:20:14.88 & +13:36:32.4 & $23.69 \pm 0.06$ & $0.50 \pm 0.12$ \\ 
3628--49 & 11:20:15.84 & +13:37:01.2 & $23.72 \pm 0.17$ & $0.34 \pm 0.21$ \\ 
3628--50 & 11:20:16.80 & +13:35:38.4 & $23.80 \pm 0.10$ & $1.44 \pm 0.12$ \\ 
~ \\ [-1.8ex] \hline 
\end{tabular*}
\end{table}

\begin{table}
\caption[ ]{Photometry and astrometry of the 50 brightest globular cluster
  candidates on the WFPC2 frames of NGC~4013. The object list is sorted on
  $V$ magnitude (brightest first).}
\label{t:phot4013}
\begin{tabular*}{8.38cm}{@{\extracolsep{\fill}}lrrcc@{}} \hline \hline
\multicolumn{3}{c}{~~} \\ [-1.8ex]  
\multicolumn{1}{c}{ID} & \multicolumn{1}{c}{RA} & \multicolumn{1}{c}{DEC} &
 $V$ & $V\!-\!I$ \\ 
  & \multicolumn{1}{c}{(J2000)} & \multicolumn{1}{c}{(J2000)} & mag & mag 
  \\ [0.5ex] \hline 
\multicolumn{3}{c}{~~} \\ [-1.8ex]  
 4013--1 & 11:58:34.32 & +43:56:24.0 & $21.30 \pm 0.01$ & $1.05\pm 0.04$ \\ 
 4013--2 & 11:58:30.96 & +43:56:34.8 & $21.78 \pm 0.02$ & $1.41\pm 0.03$ \\ 
 4013--3 & 11:58:29.52 & +43:56:02.4 & $21.88 \pm 0.02$ & $0.91\pm 0.03$ \\ 
 4013--4 & 11:58:33.36 & +43:57:18.0 & $22.12 \pm 0.02$ & $0.80\pm 0.03$ \\ 
 4013--5 & 11:58:35.28 & +43:57:46.8 & $22.20 \pm 0.01$ & $0.67\pm 0.02$ \\ 
~ & & & & \\ [-1.8ex]                                                            
 4013--6 & 11:58:29.76 & +43:56:52.8 & $22.25 \pm 0.02$ & $1.41\pm 0.03$ \\ 
 4013--7 & 11:58:32.64 & +43:56:16.8 & $22.36 \pm 0.04$ & $1.08\pm 0.06$ \\ 
 4013--8 & 11:58:30.24 & +43:57:14.4 & $22.44 \pm 0.02$ & $1.00\pm 0.03$ \\ 
 4013--9 & 11:58:31.44 & +43:56:13.2 & $22.45 \pm 0.02$ & $1.12\pm 0.05$ \\ 
4013--10 & 11:58:33.84 & +43:57:25.2 & $22.51 \pm 0.03$ & $0.78\pm 0.05$ \\ 
~ & & & & \\ [-1.8ex]                                                            
4013--11 & 11:58:28.56 & +43:56:52.8 & $22.55 \pm 0.02$ & $1.00\pm 0.04$ \\ 
4013--12 & 11:58:31.68 & +43:57:00.0 & $22.69 \pm 0.05$ & $1.22\pm 0.07$ \\ 
4013--13 & 11:58:33.36 & +43:56:45.6 & $22.74 \pm 0.02$ & $1.10\pm 0.03$ \\ 
4013--14 & 11:58:34.08 & +43:56:24.0 & $22.76 \pm 0.10$ & $0.91\pm 0.16$ \\ 
4013--15 & 11:58:25.44 & +43:57:43.2 & $22.80 \pm 0.10$ & $0.56\pm 0.11$ \\ 
~ & & & & \\ [-1.8ex]                                                            
4013--16 & 11:58:31.44 & +43:58:08.4 & $22.83 \pm 0.02$ & $0.59\pm 0.03$ \\ 
4013--17 & 11:58:30.00 & +43:57:21.6 & $22.83 \pm 0.01$ & $0.83\pm 0.03$ \\ 
4013--18 & 11:58:29.52 & +43:56:56.4 & $22.88 \pm 0.04$ & $0.96\pm 0.05$ \\ 
4013--19 & 11:58:32.40 & +43:56:34.8 & $22.89 \pm 0.04$ & $1.08\pm 0.05$ \\ 
4013--20 & 11:58:27.36 & +43:56:09.6 & $22.98 \pm 0.03$ & $0.94\pm 0.08$ \\ 
~ & & & & \\ [-1.8ex]                                                            
4013--21 & 11:58:32.88 & +43:56:27.6 & $23.02 \pm 0.04$ & $1.18\pm 0.06$ \\ 
4013--22 & 11:58:30.96 & +43:57:00.0 & $23.06 \pm 0.06$ & $1.23\pm 0.07$ \\ 
4013--23 & 11:58:33.36 & +43:55:51.6 & $23.18 \pm 0.10$ & $0.95\pm 0.13$ \\ 
4013--24 & 11:58:31.68 & +43:57:10.8 & $23.18 \pm 0.04$ & $0.71\pm 0.06$ \\ 
4013--25 & 11:58:29.76 & +43:57:10.8 & $23.29 \pm 0.08$ & $0.85\pm 0.09$ \\ 
~ & & & & \\ [-1.8ex]                                                            
4013--26 & 11:58:31.20 & +43:57:57.6 & $23.32 \pm 0.03$ & $0.77\pm 0.05$ \\ 
4013--27 & 11:58:32.88 & +43:56:34.8 & $23.33 \pm 0.05$ & $1.09\pm 0.08$ \\ 
4013--28 & 11:58:28.80 & +43:56:13.2 & $23.34 \pm 0.05$ & $1.07\pm 0.07$ \\ 
4013--29 & 11:58:32.40 & +43:57:57.6 & $23.44 \pm 0.04$ & $1.07\pm 0.04$ \\ 
4013--30 & 11:58:31.20 & +43:57:03.6 & $23.46 \pm 0.06$ & $1.01\pm 0.08$ \\ 
~ & & & & \\ [-1.8ex]                                                            
4013--31 & 11:58:32.64 & +43:56:42.0 & $23.54 \pm 0.06$ & $1.35\pm 0.08$ \\ 
4013--32 & 11:58:31.92 & +43:56:42.0 & $23.62 \pm 0.06$ & $1.36\pm 0.12$ \\ 
4013--33 & 11:58:29.52 & +43:56:49.2 & $23.66 \pm 0.06$ & $1.05\pm 0.10$ \\ 
4013--34 & 11:58:24.72 & +43:57:25.2 & $23.71 \pm 0.03$ & $0.64\pm 0.04$ \\ 
4013--35 & 11:58:30.72 & +43:57:39.6 & $23.77 \pm 0.02$ & $1.47\pm 0.06$ \\   
~ & & & & \\ [-1.8ex]                                                            
4013--36 & 11:58:33.84 & +43:57:32.4 & $23.77 \pm 0.04$ & $0.91\pm 0.05$ \\ 
4013--37 & 11:58:31.20 & +43:56:16.8 & $23.78 \pm 0.04$ & $0.95\pm 0.07$ \\ 
4013--38 & 11:58:30.72 & +43:56:34.8 & $23.81 \pm 0.07$ & $1.08\pm 0.12$ \\ 
4013--39 & 11:58:33.84 & +43:56:20.4 & $23.96 \pm 0.05$ & $1.13\pm 0.09$ \\ 
4013--40 & 11:58:28.56 & +43:57:03.6 & $23.97 \pm 0.08$ & $0.87\pm 0.11$ \\ 
~ & & & & \\ [-1.8ex]                                                            
4013--41 & 11:58:30.00 & +43:57:07.2 & $23.99 \pm 0.05$ & $0.82\pm 0.09$ \\ 
4013--42 & 11:58:32.40 & +43:57:14.4 & $24.04 \pm 0.06$ & $0.76\pm 0.09$ \\ 
4013--43 & 11:58:30.24 & +43:56:56.4 & $24.08 \pm 0.09$ & $1.08\pm 0.12$ \\ 
4013--44 & 11:58:33.12 & +43:56:38.4 & $24.10 \pm 0.08$ & $1.10\pm 0.10$ \\ 
4013--45 & 11:58:31.68 & +43:58:15.6 & $24.22 \pm 0.10$ & $1.04\pm 0.11$ \\ 
~ & & & & \\ [-1.8ex]                                                            
4013--46 & 11:58:28.80 & +43:56:52.8 & $24.25 \pm 0.08$ & $1.02\pm 0.12$ \\ 
4013--47 & 11:58:34.08 & +43:56:42.0 & $24.28 \pm 0.09$ & $0.88\pm 0.12$ \\ 
4013--48 & 11:58:33.84 & +43:56:45.6 & $24.33 \pm 0.07$ & $0.98\pm 0.13$ \\ 
4013--49 & 11:58:31.92 & +43:56:34.8 & $24.36 \pm 0.06$ & $0.99\pm 0.11$ \\ 
4013--50 & 11:58:33.84 & +43:57:28.8 & $24.43 \pm 0.05$ & $0.72\pm 0.10$ \\ 
~ \\ [-1.8ex] \hline 
\end{tabular*}
\end{table}

\begin{table}
\caption[ ]{Photometry and astrometry of the 50 brightest globular cluster
  candidates on the WFPC2 frames of NGC~4517. The object list is sorted on
  $V$ magnitude (brightest first).}
\label{t:phot4517}
\begin{tabular*}{8.38cm}{@{\extracolsep{\fill}}lrrcc@{}} \hline \hline
\multicolumn{3}{c}{~~} \\ [-1.8ex]  
\multicolumn{1}{c}{ID} & \multicolumn{1}{c}{RA} & \multicolumn{1}{c}{DEC} &
 $V$ & $V\!-\!I$ \\ 
  & \multicolumn{1}{c}{(J2000)} & \multicolumn{1}{c}{(J2000)} & mag & mag 
  \\ [0.5ex] \hline 
\multicolumn{3}{c}{~~} \\ [-1.8ex]  
4517--1  & 12:32:43.68 &   +00:06:07.2 & $21.70 \pm 0.02$ & $0.86 \pm 0.02$ \\
4517--2  & 12:32:47.52 &   +00:06:00.0 & $21.91 \pm 0.02$ & $1.24 \pm 0.02$ \\
4517--3  & 12:32:43.68 &   +00:06:32.4 & $22.05 \pm 0.03$ & $1.20 \pm 0.03$ \\
4517--4  & 12:32:42.48 &   +00:05:16.8 & $22.07 \pm 0.05$ & $0.88 \pm 0.06$ \\
4517--5  & 12:32:48.72 &   +00:05:31.2 & $22.10 \pm 0.02$ & $1.87 \pm 0.03$ \\
~ & & & & \\ [-1.8ex]                                                            
4517--6  & 12:32:46.32 &   +00:05:16.8 & $22.13 \pm 0.02$ & $1.90 \pm 0.11$ \\
4517--7  & 12:32:42.96 &   +00:06:03.6 & $22.14 \pm 0.07$ & $0.48 \pm 0.08$ \\
4517--8  & 12:32:43.44 &   +00:07:08.4 & $22.20 \pm 0.04$ & $1.22 \pm 0.06$ \\
4517--9  & 12:32:42.72 &   +00:05:38.4 & $22.21 \pm 0.02$ & $0.67 \pm 0.02$ \\
4517--10 & 12:32:46.32 &   +00:06:28.8 & $22.24 \pm 0.03$ & $0.86 \pm 0.03$ \\
~ & & & & \\ [-1.8ex]                                                            
4517--11 & 12:32:41.52 &   +00:06:25.2 & $22.33 \pm 0.21$ & $0.52 \pm 0.22$ \\
4517--12 & 12:32:44.16 &   +00:06:32.4 & $22.43 \pm 0.11$ & $1.42 \pm 0.13$ \\
4517--13 & 12:32:43.92 &   +00:06:32.4 & $22.56 \pm 0.04$ & $1.26 \pm 0.05$ \\
4517--14 & 12:32:43.20 &   +00:06:28.8 & $22.57 \pm 0.09$ & $0.33 \pm 0.12$ \\
4517--15 & 12:32:42.96 &   +00:06:21.6 & $22.89 \pm 0.06$ & $1.11 \pm 0.06$ \\
~ & & & & \\ [-1.8ex]                                                            
4517--16 & 12:32:43.68 &   +00:06:21.6 & $23.07 \pm 0.03$ & $0.98 \pm 0.05$ \\
4517--17 & 12:32:46.08 &   +00:06:28.8 & $23.08 \pm 0.13$ & $0.90 \pm 0.16$ \\
4517--18 & 12:32:44.40 &   +00:06:25.2 & $23.18 \pm 0.05$ & $0.69 \pm 0.09$ \\
4517--19 & 12:32:41.04 &   +00:04:58.8 & $23.31 \pm 0.06$ & $1.03 \pm 0.09$ \\
4517--20 & 12:32:48.48 &   +00:05:27.6 & $23.38 \pm 0.07$ & $0.55 \pm 0.08$ \\
~ & & & & \\ [-1.8ex]                                                            
4517--21 & 12:32:47.76 &   +00:05:24.0 & $23.50 \pm 0.06$ & $1.93 \pm 0.11$ \\
4517--22 & 12:32:45.12 &   +00:06:25.2 & $23.62 \pm 0.10$ & $0.63 \pm 0.13$ \\
4517--23 & 12:32:46.32 &   +00:06:36.0 & $23.64 \pm 0.10$ & $0.34 \pm 0.13$ \\
4517--24 & 12:32:45.36 &   +00:06:36.0 & $23.71 \pm 0.08$ & $0.56 \pm 0.12$ \\
4517--25 & 12:32:44.88 &   +00:06:25.2 & $23.72 \pm 0.07$ & $0.42 \pm 0.10$ \\
~ & & & & \\ [-1.8ex]                                                            
4517--26 & 12:32:43.20 &   +00:07:08.4 & $23.73 \pm 0.09$ & $0.36 \pm 0.11$ \\
4517--27 & 12:32:45.84 &   +00:06:21.6 & $23.78 \pm 0.12$ & $0.74 \pm 0.15$ \\
4517--28 & 12:32:45.36 &   +00:06:32.4 & $23.80 \pm 0.09$ & $0.64 \pm 0.11$ \\
4517--29 & 12:32:43.20 &   +00:07:12.0 & $23.84 \pm 0.07$ & $0.35 \pm 0.09$ \\
4517--30 & 12:32:45.60 &   +00:04:55.2 & $23.84 \pm 0.25$ & $1.58 \pm 0.29$ \\
~ & & & & \\ [-1.8ex]                                                            
4517--31 & 12:32:45.84 &   +00:06:28.8 & $23.85 \pm 0.10$ & $0.62 \pm 0.14$ \\
4517--32 & 12:32:45.36 &   +00:06:32.4 & $23.86 \pm 0.08$ & $0.32 \pm 0.11$ \\
4517--33 & 12:32:45.60 &   +00:06:25.2 & $23.87 \pm 0.08$ & $1.04 \pm 0.11$ \\
4517--34 & 12:32:43.44 &   +00:06:21.6 & $23.89 \pm 0.12$ & $1.88 \pm 0.14$ \\
4517--35 & 12:32:43.20 &   +00:06:32.4 & $23.92 \pm 0.07$ & $1.69 \pm 0.08$ \\
~ & & & & \\ [-1.8ex]                                                            
4517--36 & 12:32:45.60 &   +00:06:25.2 & $23.95 \pm 0.13$ & $0.36 \pm 0.17$ \\
4517--37 & 12:32:44.40 &   +00:06:25.2 & $24.02 \pm 0.08$ & $0.38 \pm 0.14$ \\
4517--38 & 12:32:42.00 &   +00:06:25.2 & $24.07 \pm 0.16$ & $1.48 \pm 0.18$ \\
4517--39 & 12:32:45.84 &   +00:06:36.0 & $24.08 \pm 0.11$ & $0.70 \pm 0.13$ \\
4517--40 & 12:32:44.40 &   +00:06:21.6 & $24.10 \pm 0.10$ & $0.58 \pm 0.13$ \\
~ & & & & \\ [-1.8ex]                                                            
4517--41 & 12:32:42.96 &   +00:06:14.4 & $24.19 \pm 0.19$ & $0.82 \pm 0.21$ \\
4517--42 & 12:32:44.40 &   +00:06:10.8 & $24.20 \pm 0.11$ & $0.90 \pm 0.15$ \\
4517--43 & 12:32:43.44 &   +00:06:25.2 & $24.23 \pm 0.19$ & $0.65 \pm 0.22$ \\
4517--44 & 12:32:45.84 &   +00:06:36.0 & $24.24 \pm 0.11$ & $0.95 \pm 0.14$ \\
4517--45 & 12:32:45.84 &   +00:06:32.4 & $24.28 \pm 0.08$ & $0.73 \pm 0.13$ \\
~ & & & & \\ [-1.8ex]                                                            
4517--46 & 12:32:45.84 &   +00:06:36.0 & $24.30 \pm 0.16$ & $0.49 \pm 0.17$ \\
4517--47 & 12:32:43.20 &   +00:06:32.4 & $24.31 \pm 0.11$ & $0.52 \pm 0.16$ \\
4517--48 & 12:32:44.40 &   +00:06:25.2 & $24.35 \pm 0.19$ & $0.53 \pm 0.21$ \\
4517--49 & 12:32:45.60 &   +00:06:32.4 & $24.41 \pm 0.12$ & $0.46 \pm 0.18$ \\
4517--50 & 12:32:45.36 &   +00:06:28.8 & $24.45 \pm 0.11$ & $0.80 \pm 0.13$ \\
~ \\ [-1.8ex] \hline 
\end{tabular*}
\end{table}

%\begin{table}
%\caption[ ]{Photometry and astrometry of the 50 brightest globular cluster
%  candidates on the WFPC2 frames of NGC~4594. The object list is sorted on
%  $V$ magnitude (brightest first).}
%\label{t:phot4594}
%\begin{tabular*}{8.38cm}{@{\extracolsep{\fill}}lrrcc@{}} \hline \hline
%\multicolumn{3}{c}{~~} \\ [-1.8ex]  
%\multicolumn{1}{c}{ID} & \multicolumn{1}{c}{RA} & \multicolumn{1}{c}{DEC} &
% $V$ & $V\!-\!I$ \\ 
%  & \multicolumn{1}{c}{(J2000)} & \multicolumn{1}{c}{(J2000)} & mag & mag 
%  \\ [0.5ex] \hline 
%\multicolumn{3}{c}{~~} \\ [-1.8ex]  
%
%~ \\ [-1.8ex] \hline 
%\end{tabular*}
%\end{table}

\begin{table}
\caption[ ]{Photometry and astrometry of the 50 brightest globular cluster
  candidates on the WFPC2 frames of IC~5176. The object list is sorted on
  $V$ magnitude (brightest first).}
\label{t:phot5176}
\begin{tabular*}{8.38cm}{@{\extracolsep{\fill}}lrrcc@{}} \hline \hline
\multicolumn{3}{c}{~~} \\ [-1.8ex]  
\multicolumn{1}{c}{ID} & \multicolumn{1}{c}{RA} & \multicolumn{1}{c}{DEC} &
 $V$ & $V\!-\!I$ \\ 
  & \multicolumn{1}{c}{(J2000)} & \multicolumn{1}{c}{(J2000)} & mag & mag 
  \\ [0.5ex] \hline 
\multicolumn{3}{c}{~~} \\ [-1.8ex]  
5176--1  & 22:14:54.96 & $-$66:50:38.4 & $21.97 \pm 0.04$ & $1.05 \pm 0.04$ \\ 
5176--2  & 22:14:43.44 & $-$66:51:03.6 & $22.18 \pm 0.06$ & $0.90 \pm 0.07$ \\   
5176--3  & 22:15:00.00 & $-$66:51:14.4 & $22.41 \pm 0.02$ & $1.82 \pm 0.02$ \\
5176--4  & 22:15:05.04 & $-$66:50:09.6 & $23.25 \pm 0.04$ & $1.18 \pm 0.05$ \\
5176--5  & 22:14:51.84 & $-$66:51:18.0 & $23.25 \pm 0.04$ & $1.13 \pm 0.04$ \\
~ & & & & \\ [-1.8ex]                                                            
5176--6  & 22:15:01.68 & $-$66:50:34.8 & $23.33 \pm 0.06$ & $1.19 \pm 0.08$ \\
5176--7  & 22:14:56.88 & $-$66:51:25.2 & $23.54 \pm 0.05$ & $1.17 \pm 0.05$ \\
5176--8  & 22:15:00.96 & $-$66:50:42.0 & $23.63 \pm 0.04$ & $1.04 \pm 0.07$ \\
5176--9  & 22:14:52.08 & $-$66:51:00.0 & $23.68 \pm 0.04$ & $1.02 \pm 0.07$ \\
5176--10 & 22:14:51.12 & $-$66:51:32.4 & $23.76 \pm 0.06$ & $1.14 \pm 0.07$ \\
~ & & & & \\ [-1.8ex]                                                            
5176--11 & 22:14:50.16 & $-$66:51:21.6 & $23.79 \pm 0.04$ & $1.07 \pm 0.06$ \\
5176--12 & 22:14:51.36 & $-$66:51:21.6 & $23.81 \pm 0.05$ & $1.01 \pm 0.06$ \\
5176--13 & 22:14:57.12 & $-$66:51:18.0 & $23.81 \pm 0.04$ & $1.17 \pm 0.05$ \\
5176--14 & 22:14:54.96 & $-$66:51:32.4 & $23.82 \pm 0.05$ & $1.42 \pm 0.06$ \\
5176--15 & 22:15:02.88 & $-$66:49:51.6 & $23.91 \pm 0.08$ & $0.73 \pm 0.10$ \\
~ & & & & \\ [-1.8ex]                                                            
5176--16 & 22:14:58.08 & $-$66:51:39.6 & $23.93 \pm 0.11$ & $0.92 \pm 0.11$ \\
5176--17 & 22:14:59.76 & $-$66:49:55.2 & $24.00 \pm 0.09$ & $0.84 \pm 0.10$ \\
5176--18 & 22:15:02.64 & $-$66:50:09.6 & $24.20 \pm 0.06$ & $0.98 \pm 0.07$ \\
5176--19 & 22:14:52.80 & $-$66:51:00.0 & $24.27 \pm 0.06$ & $1.11 \pm 0.07$ \\
5176--20 & 22:14:53.52 & $-$66:50:56.4 & $24.48 \pm 0.11$ & $1.30 \pm 0.12$ \\
~ & & & & \\ [-1.8ex]                                                            
5176--21 & 22:14:58.56 & $-$66:49:55.2 & $24.49 \pm 0.06$ & $0.83 \pm 0.08$ \\
5176--22 & 22:14:56.64 & $-$66:51:18.0 & $24.53 \pm 0.09$ & $1.09 \pm 0.10$ \\
5176--23 & 22:14:51.36 & $-$66:51:18.0 & $24.57 \pm 0.07$ & $0.89 \pm 0.10$ \\
5176--24 & 22:14:51.60 & $-$66:51:07.2 & $24.60 \pm 0.09$ & $1.05 \pm 0.11$ \\
5176--25 & 22:14:56.64 & $-$66:51:39.6 & $24.66 \pm 0.08$ & $0.78 \pm 0.09$ \\
~ & & & & \\ [-1.8ex]                                                            
5176--26 & 22:15:03.84 & $-$66:50:20.4 & $24.67 \pm 0.06$ & $1.11 \pm 0.08$ \\
5176--27 & 22:14:53.52 & $-$66:50:56.4 & $24.71 \pm 0.10$ & $0.40 \pm 0.16$ \\
5176--28 & 22:14:51.84 & $-$66:51:18.0 & $24.71 \pm 0.09$ & $0.58 \pm 0.11$ \\
5176--29 & 22:15:00.24 & $-$66:50:38.4 & $24.74 \pm 0.10$ & $1.14 \pm 0.11$ \\
5176--30 & 22:14:56.40 & $-$66:51:25.2 & $24.81 \pm 0.12$ & $0.90 \pm 0.13$ \\
~ & & & & \\ [-1.8ex]                                                            
5176--31 & 22:14:54.72 & $-$66:50:42.0 & $24.83 \pm 0.11$ & $0.87 \pm 0.13$ \\
5176--32 & 22:14:56.64 & $-$66:50:24.0 & $24.86 \pm 0.12$ & $0.95 \pm 0.14$ \\
5176--33 & 22:15:01.92 & $-$66:50:13.2 & $24.92 \pm 0.07$ & $1.28 \pm 0.11$ \\
5176--34 & 22:14:55.92 & $-$66:50:24.0 & $24.92 \pm 0.09$ & $0.90 \pm 0.11$ \\
5176--35 & 22:14:53.04 & $-$66:50:52.8 & $24.93 \pm 0.14$ & $0.46 \pm 0.24$ \\
~ & & & & \\ [-1.8ex]                                                            
5176--36 & 22:14:51.12 & $-$66:51:18.0 & $24.97 \pm 0.08$ & $0.75 \pm 0.09$ \\
5176--37 & 22:15:05.04 & $-$66:50:09.6 & $25.03 \pm 0.10$ & $1.40 \pm 0.11$ \\
5176--38 & 22:14:55.68 & $-$66:50:31.2 & $25.09 \pm 0.20$ & $1.30 \pm 0.22$ \\
5176--39 & 22:14:55.68 & $-$66:50:31.2 & $25.09 \pm 0.20$ & $1.29 \pm 0.22$ \\
5176--40 & 22:14:51.84 & $-$66:51:21.6 & $25.09 \pm 0.11$ & $1.15 \pm 0.12$ \\
~ & & & & \\ [-1.8ex]                                                            
5176--41 & 22:15:04.08 & $-$66:49:48.0 & $25.10 \pm 0.09$ & $0.37 \pm 0.12$ \\
5176--42 & 22:14:54.96 & $-$66:51:36.0 & $25.18 \pm 0.09$ & $1.37 \pm 0.11$ \\
5176--43 & 22:14:50.64 & $-$66:51:18.0 & $25.21 \pm 0.11$ & $0.78 \pm 0.13$ \\
5176--44 & 22:15:01.68 & $-$66:50:45.6 & $25.30 \pm 0.09$ & $1.13 \pm 0.11$ \\
5176--45 & 22:15:01.20 & $-$66:50:45.6 & $25.32 \pm 0.15$ & $1.09 \pm 0.18$ \\
~ & & & & \\ [-1.8ex]                                                            
5176--46 & 22:14:59.52 & $-$66:49:58.8 & $25.35 \pm 0.10$ & $0.96 \pm 0.14$ \\
5176--47 & 22:15:01.68 & $-$66:50:56.4 & $25.42 \pm 0.17$ & $1.28 \pm 0.19$ \\
5176--48 & 22:14:52.56 & $-$66:50:13.2 & $25.42 \pm 0.19$ & $1.32 \pm 0.19$ \\
5176--49 & 22:14:51.36 & $-$66:51:21.6 & $25.46 \pm 0.11$ & $1.51 \pm 0.13$ \\
5176--50 & 22:14:54.72 & $-$66:50:20.4 & $25.49 \pm 0.09$ & $1.09 \pm 0.10$ \\
~ \\ [-1.8ex] \hline 
\end{tabular*}
\end{table}

\begin{table}
\caption[ ]{Photometry and astrometry of the globular cluster
  candidates on the WFPC2 frames of NGC~7814. The object list is sorted on
  $V$ magnitude (brightest first).}
\label{t:phot7814}
\begin{tabular*}{8.38cm}{@{\extracolsep{\fill}}lrrcc@{}} \hline \hline
\multicolumn{3}{c}{~~} \\ [-1.8ex]  
\multicolumn{1}{c}{ID} & \multicolumn{1}{c}{RA} & \multicolumn{1}{c}{DEC} &
 $V$ & $V\!-\!I$ \\ 
  & \multicolumn{1}{c}{(J2000)} & \multicolumn{1}{c}{(J2000)} & mag & mag 
  \\ [0.5ex] \hline 
\multicolumn{3}{c}{~~} \\ [-1.8ex]  
7814--1  & 00:03:13.20 &   +16:08:27.6 & $21.19 \pm 0.03$ & $0.78 \pm 0.08$ \\
7814--2  & 00:03:09.84 &   +16:06:50.4 & $21.34 \pm 0.03$ & $0.61 \pm 0.03$ \\
7814--3  & 00:03:10.80 &   +16:08:02.4 & $21.43 \pm 0.15$ & $0.43 \pm 0.17$ \\
7814--4  & 00:03:08.64 &   +16:08:49.2 & $21.54 \pm 0.10$ & $0.68 \pm 0.10$ \\
7814--5  & 00:03:13.44 &   +16:08:24.0 & $22.15 \pm 0.08$ & $1.12 \pm 0.09$ \\
~ & & & & \\ [-1.8ex]                                                            
7814--6  & 00:03:12.72 &   +16:08:20.4 & $22.44 \pm 0.07$ & $0.60 \pm 0.12$ \\
7814--7  & 00:03:11.28 &   +16:07:55.2 & $22.67 \pm 0.17$ & $0.80 \pm 0.18$ \\
7814--8  & 00:03:12.24 &   +16:08:42.0 & $22.73 \pm 0.06$ & $0.30 \pm 0.13$ \\
7814--9  & 00:03:12.48 &   +16:07:26.4 & $23.03 \pm 0.04$ & $1.05 \pm 0.06$ \\
7814--10 & 00:03:10.80 &   +16:08:45.6 & $23.09 \pm 0.19$ & $0.84 \pm 0.22$ \\
~ & & & & \\ [-1.8ex]                                                            
7814--11 & 00:03:14.16 &   +16:06:50.4 & $23.27 \pm 0.10$ & $0.88 \pm 0.11$ \\
7814--12 & 00:03:13.20 &   +16:06:57.6 & $23.27 \pm 0.03$ & $1.09 \pm 0.04$ \\
7814--13 & 00:03:15.12 &   +16:08:13.2 & $23.32 \pm 0.08$ & $1.05 \pm 0.09$ \\
7814--14 & 00:03:12.96 &   +16:07:30.0 & $23.42 \pm 0.03$ & $1.72 \pm 0.14$ \\
7814--15 & 00:03:13.68 &   +16:08:24.0 & $24.12 \pm 0.14$ & $1.17 \pm 0.20$ \\
~ & & & & \\ [-1.8ex]                                                            
7814--16 & 00:03:13.44 &   +16:08:24.0 & $24.32 \pm 0.11$ & $0.76 \pm 0.16$ \\
7814--17 & 00:03:13.44 &   +16:08:34.8 & $24.63 \pm 0.20$ & $1.92 \pm 0.22$ \\
~ \\ [-1.8ex] \hline 
\end{tabular*}
\end{table}

\end{document}